\documentclass[a4paper,aps,superscriptaddress,floatfix,nofootinbib,twocolumn,longbibliography]{revtex4-1}

\usepackage{amsmath, amsthm, amssymb} % Standard math packages
\usepackage{algpseudocode} % Pseudocode package
\usepackage{comment} % Comment environment
\usepackage{graphicx}% Include figure files
\usepackage{dcolumn}% Align table columns on decimal point
\usepackage{bm}% bold math
\usepackage{hyperref}% add hypertext capabilities
\usepackage[mathlines]{lineno}% Enable numbering of text and display math
\usepackage{float}
\theoremstyle{definition}

\newcommand{\SM}{SM}

\begin{document}
%\preprint{APS/123-QED} 
% --------------------- TITLE AND ABSTRACT ----------------------
% --- Title and authorship ----

% \title{Phase transitions in optimal transportation networks}
%Title candidates
%
\title{Symmetry breaking in optimal transport networks}
%
% Phase transitions and spatial inequalities in optimal transportation networks

%spatial inequality, symmetry breaking, optimal transportation network
 
\author{Siddharth Patwardhan}
\affiliation{Center for Complex Networks and Systems Research, Luddy School of Informatics, Computing, and Engineering, Indiana University, Bloomington, Indiana 47408, USA}
\author{Marc Barthelemy}
\affiliation{Universit\'e Paris-Saclay, CNRS, CEA, Institut de Physique Th\'eorique, 91191, Gif-sur-Yvette, France}
\affiliation{Centre d’Analyse et de Math\'ematique Sociales (CNRS/EHESS) 54 Avenue de Raspail, 75006 Paris, France}
\author{\c{S}irag Erkol}
\affiliation{Center for Science of Science and Innovation, Kellogg School of Management, Northwestern University, Evanston, IL, 60208, USA}
\author{Santo Fortunato}
\affiliation{Center for Complex Networks and Systems Research, Luddy School of Informatics, Computing, and Engineering, Indiana University, Bloomington, Indiana 47408, USA}
\author{Filippo Radicchi}
\affiliation{Center for Complex Networks and Systems Research, Luddy School of Informatics, Computing, and Engineering, Indiana University, Bloomington, Indiana 47408, USA}

% ---------- Abstract ------------
\begin{abstract}

Despite its importance for practical applications, not much is known about the optimal shape of a network that connects in an efficient way a set of points. This problem can be formulated in terms of a multiplex network with a fast layer embedded in a slow one. To connect a pair of points, one can then use either the fast or slow layer, or both, with a switching cost when going from one layer to the other.  %We consider here the simpler problem of a distribution of points in the plane and ask for the fast layer network of given length that minimizes the average time to reach a central node. 
We consider here  distributions of points 
in spaces of arbitrary dimension $d$
and search for the fast-layer network of 
given 
%length 
size
that minimizes the average time to reach a central node. 
We discuss the 
$d=1$
case analytically and the 
$d>1$
case numerically, and show the existence of transitions when we vary the network 
%length, 
size,
the switching cost and/or the relative speed of the two layers. Surprisingly, there is a transition characterized by a %spatial 
symmetry breaking indicating that it is sometimes better to avoid serving a whole area in order to save on switching costs, at the expense of using more the slow layer. Our findings underscore the importance of considering switching costs while studying 
optimal network structures,
%the optimal subway structures, 
as small variations of the cost can lead to strikingly dissimilar 
results.
%optimal structures.
Finally, we discuss real-world subways and their efficiency for the cities of Atlanta, Boston, and Toronto.
%Toronto, Boston, and Atlanta. 
We find that real subways are farther away from the optimal shapes as 
%the car 
traffic congestion increases. 
\end{abstract}

\maketitle

% ---------------- SECTION INCLUSION ------------------------
% ------- Introduction ----------
%\renewcommand{\shorttitle}{Introduction}

\section{Introduction}

Networks that provide optimal transport properties 
\cite{banavar1999size,corson2010fluctuations} are of interest in many different disciplines ranging from the study of natural systems such as water transport in plants \cite{mcculloh2003water}, veination patterns in leaves \cite{katifori2010damage,mileyko2012hierarchical}, river basins \cite{rodriguez1997fractal} to the design of transportation infrastructures, either from an applied point of view \cite{laporte2015design}, or from a more mathematical perspective \cite{villani2021topics}. In particular, transportation networks evolve in time and their structure has been studied in many contexts from street networks to railways and subways \cite{xie2011evolving,roth2012long,louf2013emergence,barthelemy2022spatial,bottinelli2019efficiency,mc2020role,leite2022revealing,dahlmanns2023optimizing,bontorin2023emergence}. The evolution of transportation networks is also relevant to biological cases such as the growth of slime mould \cite{tero2010rules} or social insects \cite{latty2011structure,perna2012individual}. 

An important problem consists in designing a network from scratch or extending an existing network; this is a central subject in transportation and location science, usually known as the network design problem \cite{laporte2019introduction}. Such a problem, applied to rapid transit networks, for example, is divided into three sub-problems, which are solved numerically: location of new stations, construction of the core network connecting these stations, and location of intermediate stations on the network. From an engineering point of view, this type of problem can be solved with various optimization methods on practical cases, but the general behavior of optimal solutions is not known. From a purely mathematical point of view, there have been extensive studies of optimal networks over a given set of nodes (such as the minimum spanning tree \cite{graham1985history}, or other optimal trees \cite{barthelemy2006optimal}). Some of these problems allow for extra chosen nodes such as the Steiner tree problem \cite{hwang1992steiner}, or geometric location problems in which demand points are to be matched with supply points \cite{megiddo1984complexity}. Another example is the much-studied Monge-Kantorovich mass transportation problem \cite{rachev1998mass}, involving matching points from one distribution with points from another distribution. 

The main problem in network design is fundamentally different. We are given the density of population and we are looking for the network that minimizes some objective function involving some average time, in general (although other choices are possible, see for example \cite{laporte2015design}). In this setting, there are usually two different transport modes, a slow one representing for example cars on the road network, and a fast one representing the subway or some rapid transit network. The natural framework here is then the one of multiplex networks comprising two different transportation networks, one known while the structure of the second one is to be determined (for multiplexes in the context of optimization see for example \cite{kryven2019enhancing}). A practical realization of this problem concerns the specific case of subways (for a network analysis of subways, see for example \cite{angeloudis2006large,zhang2011networked,roth2012long,latora2002boston,lee2008statistical,derrible2009network,derrible2010characterizing,derrible2012network,leng2014evaluating,louf2014scaling}). In most large cities, a subway system has been built and later enlarged, with current total lengths varying from a few kilometers to a few hundred kilometers. The geometry of these networks, as its total length increases, varies from simple lines to more complex shapes with loops for larger networks \cite{wikipedia,aldous2019optimal}. In particular, for the largest networks, convergence to a structure with a well-connected central core and branches reaching out to suburbs has been observed \cite{roth2012long}.  

Algorithmic aspects of network design have been studied within computational geometry (e.g., \cite{okabe2009spatial} chapter 9) and location science (e.g., \cite{laporte2015design} and references therein), and some simpler problems of this type have been addressed previously. For instance, the problem of the quickest access between an area and a given point was discussed in \cite{bejan1996street,bejan1998streets}. But our specific question -- optimal network topologies as a function of population distribution and network length -- is largely an open problem. In \cite{aldous2019optimal}, 
some results were obtained in two-dimensional systems by comparing {\it a priori} defined optimal network configurations. 
First, it was shown that, if the goal is reaching a single point in the plane, then the optimal network is necessarily a tree. Second, the paper hinted at the possibility of the existence of transitions between optimal configurations when the length of the network changes. More precisely, it has been shown that as the length of the network increases resources go preferentially to radial branches and that there is a sharp transition at a critical value of the length where a loop appears.
%This study, however, doesn't prove the true optimality of those network configurations; also, it focuses only on two-dimensional systems.
%partial results were obtained in simpler cases. 

\begin{figure}[!htb]
    \centering
    \includegraphics[width=0.45\textwidth]{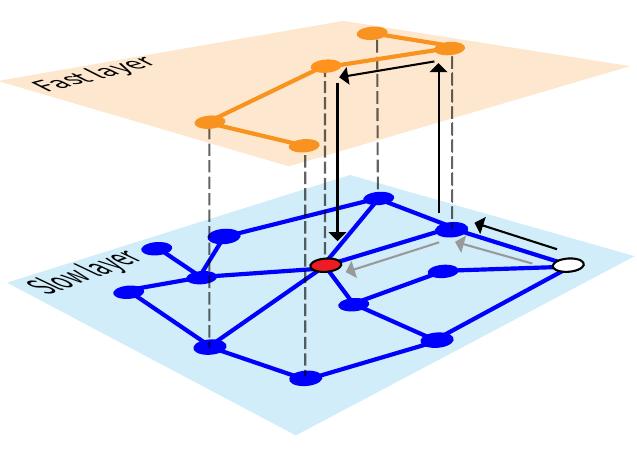}
    \caption{\textbf{Illustration of the multiplex transportation model.} In the slow layer (blue), the 
    time required to traverse an edge
    equals one; in the fast layer (orange), 
    each edge is traversed in a time 
    reduced by a factor
    $0 \leq \eta\leq 1$. Replica nodes across layers are connected by edges 
    whose transit time is
    $c \geq 0$ denoted by the dashed segments. Two possible paths connecting the white node to the center node given in red are shown. The path shown by grey arrows 
    requires a total time
    equal to $2$ as it uses only two edges in the slow layer. The second path, highlighted by black arrows, involves two changes of layer, one edge in the fast layer, and one edge in the slow layer, resulting in a total 
    transit time
    equal to  $1+ \eta + 2 c$. The second path is 
    faster
    than the first one as long as $ 1+ \eta + 2 c < 2$.} 
    \label{fig:0}
\end{figure}

In this paper, we address the problem of the quickest average time to access a central point using a multiplex framework (see Figure~\ref{fig:0}). We are given the structure of one layer, and we are allowed to build an additional layer that can facilitate quicker access to the central point. The new layer is characterized by a faster speed than the existing, slow layer; however, changes of layers incur a cost. We study the optimization problem of finding the best configuration of the fast layer on systems of arbitrary dimensions. We solve exactly the optimization problem for one-dimensional systems, showing that the optimal fast-layer configuration undergoes a sharp transition between a perfectly symmetric configuration and a fully asymmetric configuration. We numerically show that such symmetry breaking in optimal networks occurs in systems of arbitrary dimensions. We specifically focus on two-dimensional lattices and perform a systematic study of transportation systems within real cities, where we use the slow layer to model the road network and the fast layer to model the subway network. We find that real subways display network topologies compatible with the optimal ones that can be obtained using our computational framework. Differences between real and optimal networks typically arise as the ratio between subway and car speeds increases.

\section{Multiplex transportation model}

We consider a well-established network model for multimodal transportation systems (see Methods and \SM~ for details)~\cite{barthelemy2022spatial, bianconi2018multilayer}. The model comprises two network layers, namely a slow and a fast layer, each denoting a different mode of transportation, e.g., cars and subways. Each node $n$ in the slow layer has a mirror image, or replica, in the fast layer $F(n)$. For example, we can think of node $n$ in the slow layer as an intersection between roads, and of $F(n)$ in the fast layer as the subway station corresponding to that intersection. The system is such that edges in the fast layer are a subset of the replica edges of the slow layer; essentially, not all roads are mirrored by subway segments. We assume that edges in the two layers are traversed at different speeds, and without loss of generality, we parameterize the speed ratio with $0 \leq \eta \leq 1$.  Agents departing from nodes in the slow layer move along their 
quickest
path towards a specific node $o$ in the slow layer, the so-called center of the network. These 
quickest
paths can involve edges in both layers; however, each change of layer, happening between replica nodes, has a cost equal to $c \geq 0$.
See Figure~\ref{fig:0} for a schematic example. For a given configuration $\mathcal{F}$ of the fast layer, we can find the 
%shortest-path distance 
minimum-cost path
of each node $n$ in the slow layer to the center $o$. We then measure the efficiency of $\mathcal{F}$ in terms of the average 
time 
to reach the center, i.e., $\tau(\mathcal{F})$ as defined in Eq.~(\ref{eq:av_time}). The metric accounts for the fact that each node $n$ of the slow layer has associated a weight $p_n$, representing the demand of node $n$.

The goal of our modeling framework is finding the best or optimal configuration $\mathcal{F}^*$ of the fast layer, i.e., the configuration that corresponds to the minimum value of the average 
time to reach the center starting from the nodes of the slow layer. The optimization problem defined in Eq.~(\ref{eq:op}) is constrained by the
number of edges $L$ can be used to form the fast layer. Note that $L$ is interpreted as the cost of building the fast layer, hence is measured in the same units as $\tau$ and $c$.
We are interested in providing a full characterization of the topology of the optimal fast layer as a function of the parameters $\eta$ and $c$ of the multiplex transportation model. We study this optimization problem under different settings determined by the topology of the slow layer.

\section{Symmetry breaking}

\subsection{One-dimensional systems}

\begin{figure*}[!htb]
    \centering
    \includegraphics[width=0.9\textwidth]{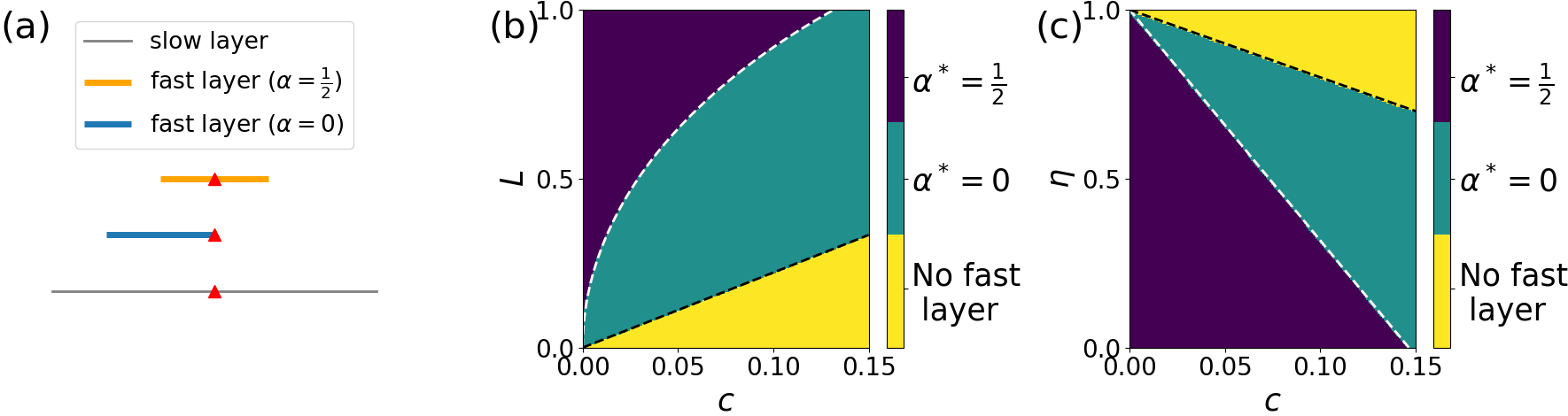}
    \caption{\textbf{Symmetry breaking in one-dimensional systems.} (a) We consider a slow layer given by a segment of length $2R$ that extends symmetrically around its center (red triangle). The fast layer is given by a segment of length $L$, with a portion of length $\alpha L$ on the right of the center and a portion of length $(1-\alpha)L$ on the left of the center, where $0 \leq \alpha \leq 1/2$. As we prove in the Methods section, two optimal configurations are possible for the fast layer: (i) a symmetric one, i.e., $\alpha = \alpha^* =1/2$, and (ii) an asymmetric one, i.e., $\alpha = \alpha^* =0$. (b) Optimal configuration of the fast layer as a function of the model parameter $L$ and the switching cost $c$. Here the ratio of the speeds of the fast and the slow layers is $\eta = 0.1$. We distinguish three regions: (i) for $L \leq r_c$, with $r_c = 2c/(1-\eta)$ as defined in Eq.~(\ref{eq:rs}) and represented by the dashed black line, the fast layer is not used; (ii) for $r_c \leq L \leq L^\dag$, with $L^\dag$ defined in Eq.~(\ref{eq:1d}) and represented by the white dashed curve, then $\alpha^* = 0$; (iii) for $L \geq L^\dag$, then $\alpha^* = 1/2$. (c) Same as in (b), but the optimal geometry of the fast layer is estimated as a function of $c$ and $\eta$ for $L=1$. The black dashed line is identified by the condition $r_c = L$; the white dashed line is given by Eq.~(\ref{eq:1d_meth}).
    }
    \label{fig:1d}
\end{figure*}

We begin our investigation by studying a one-dimensional version of our model (see Methods for details). For simplicity, the model is thought in 
continuous space. However, the calculations and results 
can be immediately generalized to a one-dimensional discrete lattice.  
The slow layer consists of a segment of length $2R$ extending symmetrically around the origin or center $o$. In the computation of the continuous version of the objective function of Eq.~(\ref{eq:av_time}), we further assume that all parts of the slow layer have equal weight.
An illustration of the system is shown in Figure \ref{fig:1d}(a). 

We remind that the problem is finding the optimal configuration of the fast layer such that the average 
%time 
cost
to reach the center from any point of the slow layer is minimum (see
Eqs.~(\ref{eq:av_time}) and~(\ref{eq:op})). 
The optimization problem is constrained by the fact that the fast layer has a fixed 
cost $L$, with  $L \leq R$.
We mathematically prove that the topology of the optimal fast layer undergoes a series of phase transitions depending on the values of the model parameters $L$, $\eta$, and $c$.

A first, trivial critical point is given by $r_c=2c/(1-\eta)$ (see also Eq.~(\ref{eq:rs}) in Methods): there is no advantage in having a fast layer with length
$L \leq r_c$, as a fast layer with such a 
cost
is not used in any 
minimum-cost
paths to the center. The optimization problem is then subject to the constraint that the fast layer should be of length $r_c \leq L \leq R$.

As we prove in the \SM, solutions to this optimization problem are given by connected segments that include the replica of the center $F(o)$. We can then parameterize the optimal fast layer by a single quantity $0\leq \alpha \leq 1/2$, such that the fast layer extends over a length $\alpha L$ to the right of 
$F(o)$ and over $(1-\alpha)L$ to the left of $F(o)$. 
We find that only two configurations for the optimal fast layer are possible: (i) a completely asymmetric configuration obtained for $\alpha = \alpha^* = 0$ 
(when $L \leq L^\dag$);
(ii) a completely symmetric configuration obtained for $\alpha = \alpha^* = 1/2$ 
(when $L \geq L^{\dag}$).
The critical 
value
$L^{\dag}$
where the transition occurs is
\begin{equation}
  L^{\dag}
  = \sqrt{4 R r_c - 2 r_c^2} \; .
  \label{eq:1d}
\end{equation}

Typical phase diagrams are displayed in Figure~\ref{fig:1d}. We clearly see
a discontinuous transition between the symmetric and the asymmetric optimal configurations as the parameters of the model vary.
This is a rather surprising 
result as it indicates that, under certain circumstances, the optimal solution is obtained by constructing a fast layer only on one side of the 
system. In other circumstances instead, a symmetric configuration is more advantageous than the asymmetric one. 

The physical intuition behind this curious behavior is as follows. Constructing a fast layer requires an initial waste of resources, as only parts of the slow layer 
%that are at a distance at least $r_c$ from the center 
whose minimum cost to reach the center is at least $r_c$
take effective advantage of the fast layer. Such an initial investment consists of building a fast layer 
such that
$L \geq r_c$ for the asymmetric case, but $L \geq 2 r_c$ for the symmetric configuration. Hence, for $r_c \leq L \leq 2 r_c$, the asymmetric configuration is trivially preferred over the symmetric one; however, the situation is not immediately inverted for $L \geq 2r_c$. As a matter of fact, any further extension of a branch of the fast layer leads to a reduction of the time to reach the center for all parts of the slow layer that are served by that branch. However, the objective function is subject to a diminishing return as the branch of the fast layer grows towards the boundary of the system. As $L$ increases, the only branch of the asymmetric configuration grows twice as fast towards the boundary than the two branches of the symmetric configuration. Thus, the initial advantage of building an asymmetric fast layer over a symmetric one is still present for $L \geq 2 r_c$, but the gap narrows as the size $L$ of the fast layer increases. The critical 
value
$L^\dag$ of Eq.~(\ref{eq:1d}) denotes the size of the fast layer when the two configurations generate identical reduction in travel time to the center and, for $L\geq L^\dag$ the symmetric configuration is preferred over the asymmetric one. The diminishing-return property of the objective function explains also why the optimal configuration for $L \geq L^\dag$ must be symmetric. If we alter in fact the symmetric configuration by reducing 
one branch 
in favor
of the other, then the increase of the objective function induced by the reduction of the one branch will be larger than the decrease of the objective function induced by the extension of the other branch. Hence, by altering the symmetric configuration, we will necessarily increase the average time to reach the center for the overall system.

Although we have mathematical support for the above interpretation only in one-dimensional systems, we believe that the general principle of symmetry breaking applies to any network regardless of the dimension of the space where the network is embedded. Indeed, in our numerical experiments we do observe symmetry breaking in the geometry of the optimal configuration of the fast layer also in systems with dimension $d >1$. We discuss these findings below.

\subsection{Two-dimensional systems}

We first extend our analysis to two-dimensional triangular lattices. The center $o$ of the slow layer is identified by the site corresponding to the geometric center of the lattice and all other nodes in the layer are identified by lattice sites at distance at most $R$ from such a center, see Methods for details. Due to the computational complexity of the optimization problem of Eq.~(\ref{eq:op}), optimal configurations of the fast layer cannot be determined exactly in this case. We rely instead on the greedy optimization strategy described in the \SM. The submodularity of the objective function of Eq.~(\ref{eq:av_time}) that we prove in the \SM~ allows us to use this algorithm to generate, in a time that roughly grows as $R^{2.7}$, approximate solutions to the optimization problem of Eq.~(\ref{eq:op}) that are at most a factor $(1-1/e) \simeq 0.63$ above the ground-truth minimum~\cite{nemhauser1978analysis}. 

\begin{figure*}[!htb]
    \centering
    \includegraphics[width=0.95\textwidth]{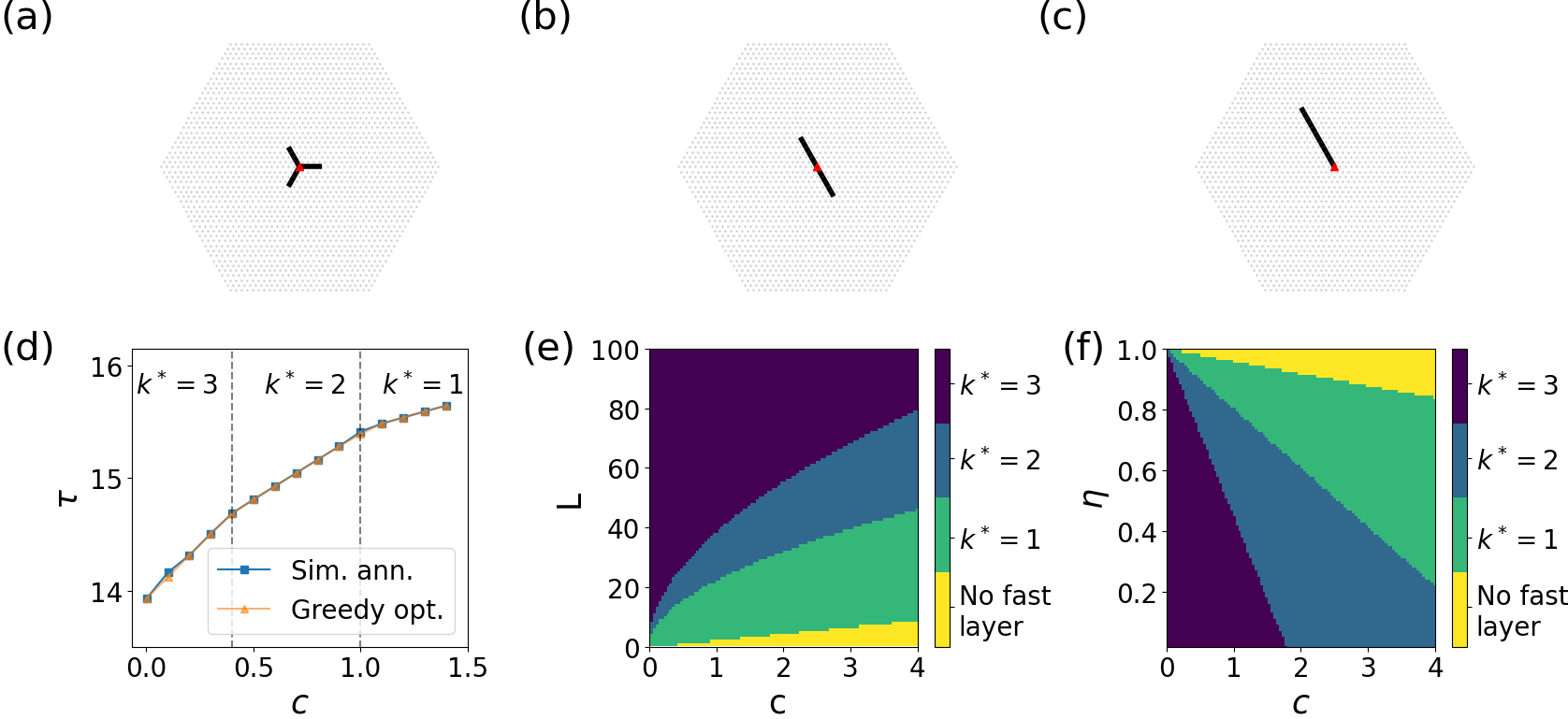}
    \caption{\textbf{Symmetry breaking in two-dimensional systems.}
    (a) Symmetric optimal configuration of the fast layer with $k^*=3$ branches. Here $R=25$ and $L=12$.
    (b) Symmetric optimal configuration of the fast layer with $k^*=2$ branches obtained for $R=25$ and $L=12$
    (c) Asymmetric optimal configuration of the fast layer with $k^*=1$ branch valid for $R=25$ and $L=12$. (d) Average time to the center, i.e., Eq.~(\ref{eq:av_time}) associated with the optimal fast-layer configuration as a function of $c$. Here, $R = 25$, $L=12$ and $\eta=0.1$. We compare solutions obtained using greedy and simulated-annealing optimization. 
    The vertical dashed lines correspond to the values of $c$ where we observe a change in the topology of the optimal fast layer. (e) Number of branches characterizing the topology of the optimal fast layer as a function of $L$ and $c$. Here, $R=100$ and $\eta=0.1$. (f) Same as in (e), but as a function of $\eta$ and $c$, with $L=50$.
    }
    \label{fig:1}
\end{figure*}

Typical solutions obtained using greedy optimization are displayed in Figures~\ref{fig:1}a, b, and c. Here, we assume that the weight associated with each node $n$ of the slow layer is $p_n=$const. 
As discussed in the Methods section and proved in the \SM, optimal configurations of the fast layer are given by trees with at least one edge incident to $F(o)$. However, depending on the choice of the model parameters $L$, $\eta$ and $c$, different optimal configurations may emerge. As for the one-dimensional case, also here optimal configurations of the fast layer appear to be characterized by branches of similar length, so that different optimal configurations can be distinguished by simply counting the number of such branches, namely $k^*$ as defined in Eq.~(\ref{eq:geo_param}). We do observe $k^*=1, 2$ and $3$ in Figures~\ref{fig:1}a, b and c, respectively. Please note that this simple characterization of the geometry of the fast layer is valid only in the regime $L < R$. For larger sizes of the fast layer, the geometry of the optimal configurations becomes much richer and requires additional order parameters to be described; in this paper, we only consider phase transitions concerning fast layers whose size is much smaller than the one of the slow layer.

Typical phase diagrams are shown in Figures~\ref{fig:1}d, e, and f. 
 As for the one-dimensional continuous model, we observe that a fully asymmetric configuration emerges for large $c$ values, see Figures~\ref{fig:1}d, e, and f; as $L$ increases, we observe transitions towards larger number of branches, see Figure~\ref{fig:1}f. In Figure\ref{fig:2}d, we validate the goodness of the solutions obtained via greedy optimization by comparing them with solutions obtained via simulated-annealing optimization (see \SM~for details). 

In the \SM, we consider a continuous-space approximation of the two-dimensional lattice. The results of our analysis are qualitatively similar to those valid for the discrete lattice, with the only caveat that optimal configurations of the fast layer can be characterized by an unbounded number of branches. In the \SM, we also consider two-dimensional lattices where the weight $p_n$ is an exponentially decreasing function of the lattice distance of node $n$ to the center $o$. Results are qualitatively similar to those reported in Figure~\ref{fig:1} in that setting too.

\subsection{Higher-dimensional systems}

Our findings on the breaking of the symmetry of the optimal fast layer generalize also to infinite-dimensional systems.

In the \SM, we consider a continuous-space approximation of a star-like system where the slow layer is given by 
an arbitrary number $q$ of segments intersecting in a single point and extending symmetrically around this central point. We 
can prove analytically that the only allowed solutions to our optimization problem are given by fast layers consisting of $1 \leq n^* \leq q$ branches of identical length $L/n^*$. 
%Depending on the parameters of the model, we can observe $0 \leq n^* \leq q$. 
As in the case of the one- and two-dimensional systems, also for the star-like system we observe that $n^* = 1$ for sufficiently large $c$ values, and that $n^*$ grows as $L$ increases.

The same qualitative behavior is also observed in numerical simulations on one instance of the Erd\H{o}s-R\'enyi model
with $N =1,000$ nodes and average degree $\langle k \rangle =4$. For simplicity, in our simulations, we select the node with the largest degree $k_{\max}=11$ as the center of the slow layer. We then determine the optimal configuration of the fast layer via greedy optimization. We observe transitions between configurations of the optimal fast layer with variable number of branches $0 \leq k^* \leq k_{\max}$ depending on the choice of the model parameters. Results of these simulations are reported in the \SM.

\section{Real-world cities}

In this section, we study the properties of the subway systems in Atlanta, Boston, and Toronto under the lens of our framework.
We choose monocentric cities, fairly isolated from other major urban centers, with a tree-like subway structure. 
%The intersection of the subway lines helps us determine the center of the city. Moreover, w
%We choose cities that 
%Also, these cities are fairly isolated from other major urban centers.
%since these centers may affect the subway structure around the city. 
We identify the intersection point of the real subway lines in all three cases as the city center. We see that 
%this is a good assumption as the 
this
point corresponds to the downtown area in the three cities. 

\begin{figure*}[!htb]
    \centering
    \includegraphics[width=0.95\textwidth]{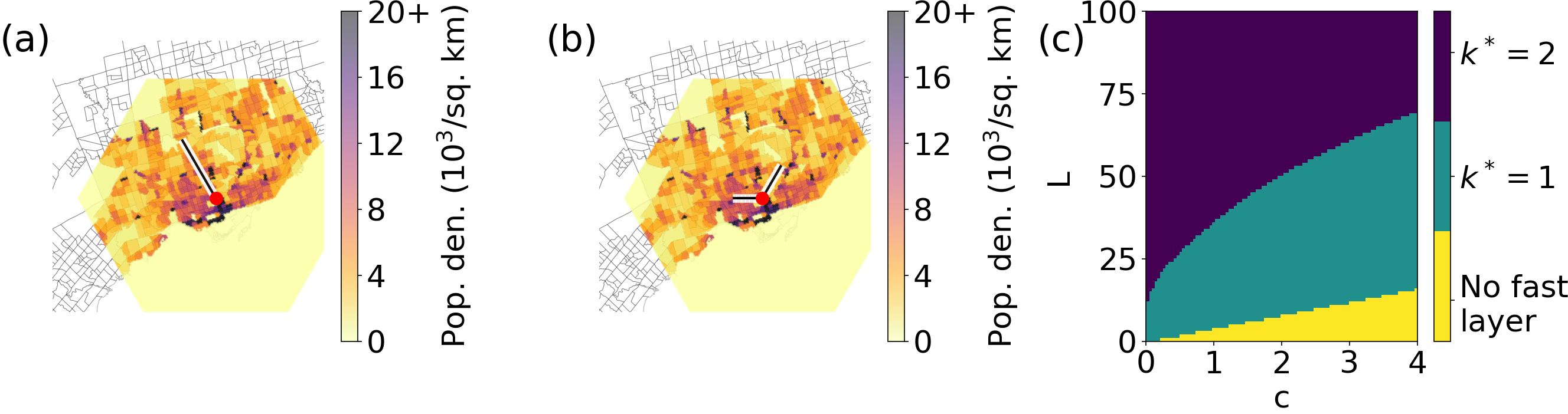}
    \caption{\textbf{Symmetry breaking in the transportation networks of real cities.} (a) We consider the city of Toronto. We construct the slow layer of the system using a triangular lattice radius $R=100$; for the fast layer, we impose $L=50$. The color map shows the population density associated with the lattice points; the gray lines represent census-tract boundaries. The red circle represents the center. We show the optimal configuration of the fast layer with $k^*=1$ branch. 
    (b) Same as in (a), but we show the solution with $k^*=2$. (c) 
    Phase diagram displaying the value of $k^*$ as a function of model parameters $L$ and $c$. Here $\eta = 0.5$. 
    The yellow region denotes $L \leq r_c$.
    }
    \label{fig:2}
\end{figure*}

First, we incorporate real population data in our model~\cite{US21Census, Canada21Census}. 
We rely on a two-dimensional triangular lattice multiplex model; we use the population data and the appropriate coordinates reference systems to impose the triangular lattice structure onto the city landscape. Details on the data and modeling of the city population distribution can be found in the Methods section.
We denote all quantities relative to the real physical system using the same notation as for the multiplex model, but we add a tilde on top of the corresponding symbol. For example, $R$ indicates the radius of the lattice model, and $\tilde{R}$ denotes the radius of the city.
Overlaying a city on top of the triangular lattice allows us to associate a weight $\tilde{p}_n$ to each node $n$ in the slow layer that reflects the real population density within the city. We use those weights in the objective function of Eq.~(\ref{eq:av_time_real}), and then take advantage of the greedy algorithm to obtain approximate solutions to the optimization problem of Eq.~(\ref{eq:op}). Similar to the previous sections, we obtain two classes of optimal fast-layer configurations for all the considered parameters. Results for the city of Toronto are displayed in Figure~\ref{fig:2}, where we see that optimal configurations comprise $k^*=1$ (Figure~\ref{fig:2}a) or $k^*=2$ (Figure~\ref{fig:2}b) branches. Similar results are valid for Atlanta and Boston, where we observe optimal configurations with $k^* \leq 3$ branches (see \SM). For $k^*> 1$, we note that the branches have no identical length; this is caused by the fact that the weight associated with the various nodes of the system is not constant. A typical phase diagram for Toronto is displayed in Figure~\ref{fig:2}c, where we fix $\eta = 0.5$, but vary $L$ and $c$.  The diagram is qualitatively similar to the one of Figure \ref{fig:1}e. For fixed $c$, $k^*$ increases as $L$ grows; however, for fixed $L$, $k^*$ decreases as $c$ grows. The values of the parameters where the transitions between the various phases emerge differ from those of Figure~\ref{fig:1}e; this is due to the non-homogeneous density of the population used in the model of the city. Similar results for Atlanta and Boston can be found in the \SM.

\begin{figure*}[!htb]
    \centering
    \includegraphics[width=0.95\textwidth]{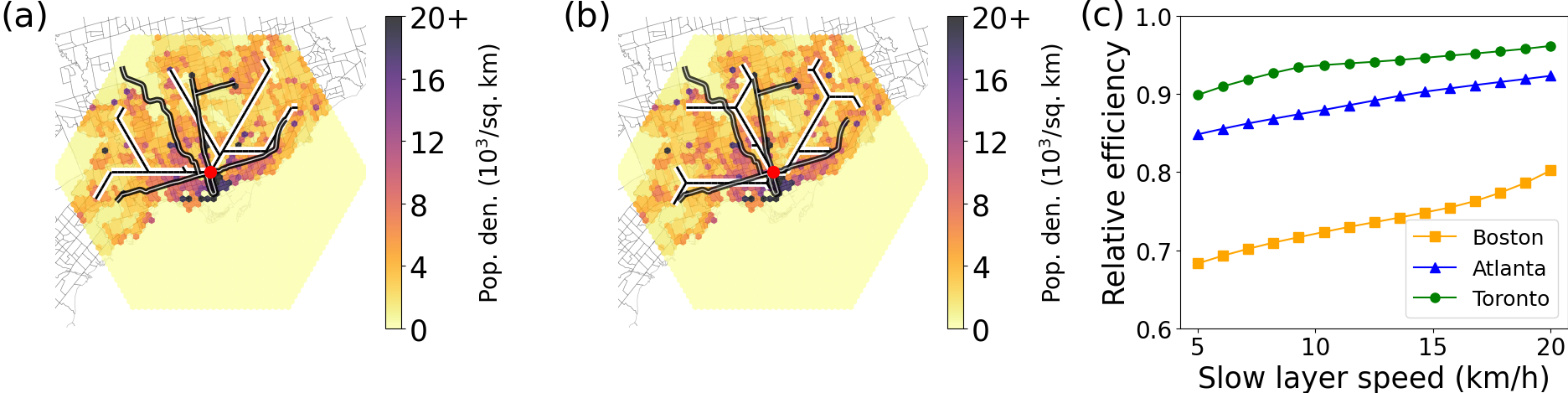}
    \caption{\textbf{Assessing the optimality of the transportation networks of real cities.} 
    (a) We compare the subway network generated with our optimization framework (white curves) with the real subway system (black curves) in the city of Toronto. The optimized configuration is obtained by setting $R=25$, $\eta=0.5$, and $c=1.25$ in the lattice model. These choices correspond to setting, in the physical system, the speed of the slow and fast layers respectively to $\tilde{v}_s = 20$ km/h and $\tilde{v}_f = 40$ km/h, and the switching time between layers to $3$ mins. (b) Same as in (a), but for $\tilde{v}_s = 5$ km/h. We also set $c=0.3125$ in the lattice model so that the switching time in the physical system still equals $3$ mins. (c) The efficiency of the real subway systems relative to the optimal configurations as a function of the speed of the slow layer $\tilde{v}_s$.
    Relative efficiency is given by the ratio between the values of the objective function estimated via Eq.~(\ref{eq:av_time_real}) for the optimized and the real configurations of the fast layer. 
    As we vary $\tilde{v}_s$, we change also the value of the parameter $c$ in the multiplex model so that the switching time in the physical system is equal to $3$ mins, see Methods for details.}
    \label{fig:3}
\end{figure*}

Next, we perform a direct comparison between the real subway lines and the optimal fast-layer configurations obtained using our computational framework. 
To this end, we calibrate the model's parameters $L$ and $R$ such that the number of subway stations in the real system is comparable with the one in the model. Results of this analysis are reported in Figure~\ref{fig:3} for Toronto and in the \SM~ for Atlanta and Boston.
We first note that the optimal fast-layer networks display additional ramifications. This is due to the fact that $L > R$ in this experimental setting. Second and more important, we note that there is an overall good overlap between the real subways and those obtained under the framework. This is true regardless of the specific choice of the model parameters (Figures~\ref{fig:3}a and b). We quantify the efficiency of the real subway systems relative to the optimal configurations by measuring the ratio of the corresponding values of the objective function of Eq.~(\ref{eq:av_time_real}), see Figure~\ref{fig:3}c. Here, we keep the speed of the fast layer invariant as $\tilde{v}_f = 40$ km/h, and vary the speed of the slow layer $\tilde{v}_s$. This corresponds to effectively varying the value of the model parameter $\eta$.
The real subway system appears less efficient than the optimized one in congested situations when $\tilde{v}_s$ is small. However, it gets close to optimality as the speeds in the slow layer grows towards the value of the speed of the fast layer.

\section{Discussion} 

Location science and network design focused on practical aspects of network optimization. Even if operational research is successful in designing minimal cost solutions, the theoretical question of the optimal network topology is largely open. In addition, and as suspected in previous studies, we showed that these optimal networks experience a transition between different shapes when the total length or the switching cost increase: small variations of the cost can lead to strikingly dissimilar optimal structures. In particular, there is a transition characterized by a symmetry breaking leading to spatial inequalities. Such a phenomenon results from the interplay between switching cost and the absence of the network and shows that an optimal solution is not necessarily isotropic. Although such phenomenon is known to happen in various economic instances \cite{fujita2001spatial}, it is the first time that it is exhibited in an optimal network context. These results also underscore the importance of considering switching costs and the cost associated with the slow layer (typically car traffic) when studying the optimal subway structures. A better theoretical understanding of these optimal shapes could certainly be helpful for practical applications and the identification of critical parameters. Further studies are however needed in order to explore in more depth these transitions. Also, we focused here on the monocentric case where we minimize the average distance to reach a central node. Large cities are however polycentric and the structure of flows is far more complex. Preliminary results suggest that here also there are transitions between different optimal shapes, but this point certainly deserves further studies. Finally, the more difficult problem of minimizing the average time needed to connect any pair of points is even more open. In this case, the optimal network can have loops and is computationally more demanding. Although we also expect transitions, our understanding of this case is at the beginning.

\section{Methods}
\subsection{Multiplex transportation model}

We consider a multiplex network 
composed of a slow layer and a fast layer (see Figure~\ref{fig:0}).
We denote with $\mathcal{G}$ the set of nodes in the slow layer, and with
$\mathcal{S}$ the set of its edges; $\mathcal{H}$ and $\mathcal{F}$ are respectively the set of nodes and edges in the fast layer.
Both layers contain $N$ nodes; each node $n$ in the slow layer has a one-to-one correspondence with a node $F(n)$ in the fast layer. Each edge $(F(n), F(m))$ in the fast layer has a replica edge $(n,m)$ in the slow layer, 
but the vice versa is not necessarily true. The 
%length of
transit time of
each edge in the slow layer equals one, whereas 
%the length associated with the 
time required to traverse
edges of the fast layer is 
%equal to 
reduced by a factor
$0 \leq \eta \leq 1$. Replica nodes are connected to each other by edges 
%of length 
whose transit time is
$c \geq 0$. 
Please note that in this mathematical framework entities have no physical meaning, thus we can interchange the notions of the length of an edge with that of the time required to traverse it, and simply refer to them with the generic term cost.

When considered in isolation, the slow layer forms a single connected component, whereas the fast layer is not necessarily connected. The connectedness of the slow layer implies, however, that in the overall system, composed of the interconnected slow and fast layers, there exists at least a path connecting any pair of nodes.  The 
%length 
cost
of a path in the network is given by the sum of the 
%lengths 
costs
of all edges that compose the path. The 
%shortest
minimum-cost
path between two nodes can either use edges in the slow layer only or take advantage of some of the edges in the fast layer (see Figure~\ref{fig:0}). In particular, the path $n \to F(n) \to F(m) \to \ldots \to F(r) \to F(s) \to s$ composed of $\ell$ edges in the fast layer only is preferred to its replica path $n \to m \to \ldots \to r \to s$ whenever $\ell$ is larger than 
%the critical length scale of the multiplex model, i.e.,
\begin{equation}
r_c = \frac{2 c}{ 1 - \eta} \; .
    \label{eq:rs}
\end{equation}

%\section{Optimization of the fast layer}

We identify a special node $o$ in the slow layer of the network, 
i.e., the center of the network. 
%The label of the center is $o$. 
We denote with $d_n$ the 
%length of the shortest path 
cost of the fastest path
of the generic 
node $n$ to $o$. Also, we assume that each node $n$ in the slow layer has an associated weight $p_n \geq 0$. We then define the weighted average 
%distance 
cost
to the center as
\begin{equation}
    \tau(\mathcal{F}) = \frac{\sum_{n \in \mathcal{G}} \, d_n \, p_n}{\sum_{n \in \mathcal{G}} \, p_n}  \; .
    \label{eq:av_time}
\end{equation}

We stress that the above function is computed over all nodes in the slow layer only, but eventual 
%shortest 
minimum-cost
paths can take advantage of edges in the fast layer. Clearly, $\tau$ depends on the various parameters of the model. In Eq.~(\ref{eq:av_time}), we explicit, on purpose, only the dependence of $\tau$ on the fast layer $\mathcal{F}$ as this is the primary object of our investigation. We consider in fact the optimization problem aimed at finding the best set of edges in the fast layer able to minimize the objective function of Eq.~(\ref{eq:av_time}). The minimization is constrained by the number of edges $L$ that are in the fast layer, with $L$ still measured in the same units of costs as $\tau$ and $c$. Specifically, we aim at solving

\begin{equation}
    \mathcal{F}^* = \arg \min_{|\mathcal{F}|=L} \, \tau(\mathcal{F}) \;, 
    \label{eq:op}
\end{equation}

where we indicated with $|\mathcal{F}|$ the number of edges in $\mathcal{F}$.

%\subsection{Optimization techniques}

Finding the exact solution to the optimization problem of Eq.~(\ref{eq:op}) is computationally infeasible as it requires a brute-force search over all possible ${|\mathcal{S}| \choose L}$ configurations of the fast layer. In the SM, we prove, however, that: (i) the optimal configuration of the fast layer is a connected tree with at least one edge incident to $F(o)$, i.e., the replica node of the center; (ii) the objective function of Eq.~(\ref{eq:av_time}) is a decreasing and submodular function. The relevance of property (i) is two-fold: first, it allows us to dramatically reduce  the number of suitable solutions for the optimization problem of Eq.~(\ref{eq:op}); second, it permits us to meaningfully describe the geometry
of the optimal fast layer in terms of the number of branches departing from the replica node of the center, i.e.,
\begin{equation}
    k^* = \sum_{(F(n), F(m)) \in \mathcal{F}^*} \, \left[ \delta_{F(o), F(m)} + \delta_{F(n), F(o)} \right] \; ,
    \label{eq:geo_param}
\end{equation}
where $\delta_{x,y} = 1$ if $x=y$ and $\delta_{x,y} = 0$ otherwise.
Property (ii) allows us to leverage a greedy optimization scheme to generate approximate solutions to the optimization problem of Eq.~(\ref{eq:op}) that are at most a fraction $(1 - 1/e)$ above the ground-truth minimum~\cite{nemhauser1978analysis}. In the construction of greedy solutions, we start from an empty set of edges in the fast layer, and we add one edge at a time. The edge that is added is the one corresponding to the best choice that can be made given the current set of edges in the fast layer. In the SM, we further describe how solutions obtained via greedy optimization can be further refined to get better approximations for the optimization problem of Eq.~(\ref{eq:op}); the quasi-optimality of our greedy solutions is validated by comparing them to those obtained via simulated-annealing optimization (see Figure~\ref{fig:1} and SM for details).

\subsection{Two-dimensional triangular lattices}
The slow layer used in the definition of the multiplex transportation model can be represented by any connected network. In the SM for example, we report on results obtained for a slow layer given by an instance of an Erd\H{o}s-R\'enyi model. 

The vast majority of the results reported in this paper are obtained for slow layers derived from triangular lattices. The coordinates of all sites of the lattice are in the form $(a+\frac{b}{2}, \frac{\sqrt{3}b}{2})$ for integer values of $a$ and $b$ such that $|a|+|b|\leq R$, where $R$ is the radius of the triangular lattice. The boundary conditions give the system a hexagonal shape. The sites of the lattice are the nodes of the slow layer; each pair of nodes in the slow layer is connected if the corresponding sites are at distance one in the triangular lattice; we identify the center of the network as the site with coordinates $(0,0)$, i.e., $a=b=0$.

\subsection{Real-world cities}

In this section, we describe how we model the transportation system of a real city. Our framework relies on the use of a multiplex network formed of two discrete triangular lattices, one used to describe slow transportation (e.g., cars) and the other used to model fast transportation (e.g., subways). As we are referring to a real physical system, all quantities that describe properties of the multiplex transportation model have an associated physical dimension. For simplicity, we still rely on the same notation as in the previous sections, however, we add a tilde on the top of the symbols to make clear that the notation is used to indicate physical quantities. For example, we use $\tilde{R}$ to denote the city radius measured in units of length and distinguish it from $R$ which serves to indicate the radius of the triangular lattice measured in dimensionless lattice units.

\subsubsection{Data}
We obtain the population density at the census-tract level for Boston and Atlanta from the 2021 American Census Survey~\cite{US21Census}, and for Toronto from the 2021 Canadian Census of Population~\cite{Canada21Census}.  The census data contains the number of individuals residing in relatively small geographic regions, i.e., census tracts, used for statistical purposes by national statistical agencies. Census tracts typically consist of $2,500$ to $8,000$ individuals. 

The data on the four metro lines and their stations in Boston is obtained from the Massachusetts Bureau of Geographic Information (MassGIS) website~\cite{massgis}. Similar data for the four metro lines in Atlanta is obtained from the Atlanta Regional Commission: Open Data website~\cite{marta}. Finally, the data for the three subway lines and their stations in Toronto is made available by the Toronto Transit Commission at the City of Toronto Open Data website~\cite{TTCSubway}. The data on the subway lines is available as shapefiles. The arcs for the rail lines are given by sets of points in the coordinate reference systems (CRS) applicable to the geographic location. The CRSs used for Atlanta, Boston, and Toronto are EPSG:2239, EPSG:26986, and, EPSG:2952, respectively. Similarly, the location of the stations is given by points in the appropriate CRS. The Atlanta, Boston, and Toronto subway systems total $\tilde{L} = 77$ km, $\tilde{L} =109.6$ km, and $\tilde{L} = 69.6$ km of rail lines and $\tilde{n}_s = 38$, $\tilde{n}_s = 125$, and $\tilde{n}_s = 75$ stations, respectively. The average distance between the stations is $2.07$ km, $0.88$ km, and $0.94$ km for Atlanta, Boston, and Toronto, respectively. 

We choose a city radius $\tilde{R}$ such that all stations are contained in the circle of radius $\tilde{R}$ around the center. We find the appropriate choice for this radius to be $\tilde{R}=25$ km in Atlanta and $\tilde{R}=20$ km in Boston and Toronto.
We assume that everything that lies inside this circular area constitutes the city.

\subsubsection{Multiplex transportation model for cities}

We create a lattice model of the transportation system in a city by overlaying a triangular lattice of radius $R$ on top of the circle of radius $\tilde{R}$. Please note that we use $R=100$ in all figures except for Figure~\ref{fig:3} where we use $R=25$. 
The operation requires matching locations of the city that are given by points in continuous space to lattice sites. To this end, we fix the location of the city center as the one of the center $o$ of the triangular lattice. Since the lattice covers the entire circular area of the city, the physical distance between neighboring sites on the lattice is $\tilde{w}_{n,m} = \tilde{R} /R$. The choice of $R$, for given $\tilde{R}$, determines the granularity of the lattice mesh overlaid on the city landscape. For instance, $R=100$ and $\tilde{R}=25$ km give us neighboring lattice sites $n$ and $m$ at distance $\tilde{w}_{n,m} = 0.25$ km, whereas $R=25$ and $\tilde{R}=25$ km give $\tilde{w}_{n,m} = 1$ km. Note that the physical distance between nodes $n$ and $m$ in the two layers is the same as the physical distance between their replica nodes $F(n)$ and $F(m)$ in the fast layer, i.e., for all $(F(n),F(m)) \in \mathcal{F}$  $\tilde{w}_{F(n),F(m)} = \tilde{w}_{n,m}$. The distance between the replica nodes $n$ and $F(n)$ is a parameter of the model $\tilde{w}_{n, F(n)} = \tilde{\ell}$.

The weight $\tilde{p}_n$ associated with node $n$ in the slow layer is given by the population density of the associated census tract containing the site.  Note that the size of census tracts varies significantly as the population density varies. Consequently, depending on the choice of $R$ and $\tilde{R}$, we may have no lattice sites in very small census tracts. We find that this issue can be resolved by choosing $R=100$ for the values of $\tilde{R}$ indicated above. 

We assume that the travel speed  $\tilde{v}_s$ on the slow layer is between $ 5 \, km/h$ and $20 \, km/h$, and the speed on the fast layer is $\tilde{v}_f =40 \, km/h$. These are both realistic (ranges of) values for the travel speeds of cars and subways, respectively. Clearly, we have $\eta = \tilde{v}_s/\tilde{v}_f$. We note that the time required to traverse the edge $(n,m) \in \mathcal{S}$  is $\tilde{t}_{n,m} =  \tilde{w}_{n,m}/\tilde{v}_s $, whereas the time required to traverse the edge $(F(n),F(m)) \in \mathcal{F}$ is  $\tilde{t}_{F(n),F(m)} = \eta \, \tilde{t}_{n,m}$. For example, if  $R=25$, $\tilde{R}=25$ km, $\tilde{v}_s  = 20$ km/h, and $\tilde{v}_f = 40$ km/h, we have $\tilde{t}_{n,m} = \frac{1}{20}$ hours or $3$ minutes, and $\tilde{w}_{F(n),F(m)}=\frac{1}{40}$ hours or $1.5$ minutes. Finally, we assume that a change of layers occurs also at speed $\tilde{v}_{s}$, meaning that the time required to switch layers is  $\tilde{t} = \tilde{\ell}/\tilde{v}_{s}$.

We denote with $\tilde{t}_n$ the time required to reach the center $o$ from node $n$. This is given by the time corresponding to the fastest path connecting the two nodes. We finally rewrite Eq.~(\ref{eq:av_time}) as
\begin{equation}
\tilde{\tau}(\mathcal{F}) = \frac{\sum_{n \in \mathcal{G}} \, \tilde{t}_n \, \tilde{p}_n}{\sum_{n \in \mathcal{G}} \, \tilde{p}_n}  \; ,
    \label{eq:av_time_real}
\end{equation}
and use this expression while solving the optimization problem of Eq.~(\ref{eq:op}).

\subsubsection{Estimating the average time to the center for real cities}

Next, we explain the modeling framework used to obtain the results for the real subway systems in Figure \ref{fig:3}.  The slow layer is modeled identically as described above. We use $\tilde{n}_s$ stations as the nodes of the fast layer. Connections between stations are given by actual railways, with length learned directly from the data. For each station, we identify its replica in the slow layer as the node corresponding to the closest (in terms of geographical distance) site of the triangular lattice. 

To properly compare the objective function of Eq.~(\ref{eq:av_time_real}) for the real city against the one obtained after greedy optimization, we set $R=25$ when constructing the triangular lattice. This allows us to obtain comparable numbers of subway stations between the 
real cities and the synthetic ones. We respectively have $L = 77, 132$, and $87$ for the synthetic versions of Atlanta, Boston, and Toronto.

\subsection{Continuous-space approximation for one-dimensional lattices}
To ease analytical calculations, we adopt a continuous-space approximation for a multiplex transportation model where the slow layer is given by a one-dimensional lattice where the weight associated to each node $n$ in the slow layer is $p_n=$const. Under the continuous-space approximation, the slow layer has the center located in the origin, and is formed of two segments of length $R$ extending symmetrically to the left and the right of the center (see Figure~\ref{fig:1d}a). The fast layer extends to the right of the origin with a segment of length $\alpha L$ and to the left with a segment of length $(1-\alpha)L$, where $0 \leq \alpha \leq 1/2$ is a tunable parameter and $L \leq R$ is the total length of the fast layer. The goal of our calculation is to find the value $\alpha^*$ corresponding to the optimal configuration of the fast layer, i.e., the solution of the continuous-space approximation of the optimization problem of Eq.~(\ref{eq:op}).

Consider first the right side of the fast layer which is serving the right side of the slow layer. The objective function relative to the right side of the system is
 \begin{equation}
  \tau_{\text{right}}(\alpha) =
  \left\{
    \begin{array}{ll}
      R^2/2 & \textrm{ if } 0 \leq \alpha \leq r_c/L
              \\
      C + (1- \eta)  L \left[ \alpha^2 L/2 - \alpha R \right] & \textrm{oth.}
    \end{array}
    \right.
      \;, 
      \label{eq:1da}    
    \end{equation}
    where
\begin{equation}
C = \frac{1}{2} (1-\eta )r_c^2 + 2c(R-r_c) +\frac{1}{2}R^2 \; .
\label{eq:const}    
\end{equation}
If $\alpha L < r_c$, the fast layer does not serve any portion of the slow layer, thus $\tau_{\text{right}}(\alpha) = \int_0^R \, dx \, x = R^2/2$. If $\alpha L \geq  r_c$, we need to solve the integral $\tau_{\text{right}}(\alpha) = \int_{0}^{r_c} \, dx \, x^2 + \int_{r_c}^{\alpha L} \, dx \, \left( 2c + \eta x \right) + \int_{\alpha L}^R \, dx \, \left( 2c + \eta \alpha L + x \right)$, leading to the second case of Eq.~(\ref{eq:1da}). We note that the term $C$ appearing in Eq.(\ref{eq:const}) does not depend on $\alpha$, but only on $r_c$ and $R$. 

For the left portion of the fast layer, we simply have $\tau_{\text{left}}(\alpha) = \tau_{\text{right}}(1 - \alpha)$. For the entire system, the objective function reads $\tau(\alpha) = \tau_{\text{right}}(\alpha) + \tau_{\text{left}}(\alpha)$.

We now distinguish two cases: (i) $r_c/L \leq 1/2$ and (ii) $r_c/L \geq 1/2$.
In case (i), we can write:
\begin{widetext}
\begin{equation}
  \tau(\alpha) =
\left\{
    \begin{array}{ll}
      R^2/2 + C + (1- \eta) L \left[ (1-\alpha)^2 L/2 - (1-\alpha) R \right] & \textrm{ if }  0 \leq \alpha \leq r_c/L
      \\
      2 C + (1- \eta) L \left[ ( (1-\alpha)^2 + \alpha^2) L/2 - R
      \right] & \textrm{ if } r_c/L \leq \alpha \leq 1/2  %1 -r_c/L
      %\\
      %R^2/2 C + (1- \eta) \alpha L \left[ \alpha L/2 - R \right] &  \alpha \geq 1 - r_c/L
      \end{array}
    \right.
    \; .
    \label{eq:1db}
  \end{equation}
\end{widetext}
 thus,
  \[
  \small
  \frac{d \tau(\alpha)}{d \alpha} =
\left\{
  \begin{array}{ll}
     (1- \eta)  L \left( R - (1- \alpha) L \right) \geq 0 & \textrm{ if }  0 \leq \alpha \leq r_c/L
      \\
      (1- \eta) L \left[  - L - R \right] \leq 0 &  \textrm{ if }  r_c/L \leq \alpha \leq 1/2 %1 -r_c/L
      \end{array}
    \right.
    \; .
    \normalsize
  \]
  We see therefore that the function reaches its maximum at $\alpha = r_c/L$, and displays its 
  minimum value either in $\alpha = 0$ or $\alpha = 1/2$.
  To determine where the minimum of the objective function of Eq.~(\ref{eq:1db}) is obtained, we need to solve the equation $\tau(\alpha=0) = \tau(\alpha=1/2)$.
  After some simple calculations, we arrive to 
  \begin{equation}
  %r_*
  r^{\dag}
  =R\left[1-\sqrt{1-\frac{1}{2}\left(\frac{L}{R}\right)^2}\right] \; .
  \label{eq:1d_meth}
\end{equation}
  %Eq.~(\ref{eq:1d}).
  %\begin{equation}
  %  \tau(0) = R^2/2 + C + (1- \eta) L \left(  L/2 - R \right) = 2 C + (1- \eta) L \left[  L/4 - R
  %    \right] = \tau (1/2)
  %  \end{equation}
  %   \begin{equation}
  %    R^2/2 - C + (1 - \eta) L^2/4 = 0
  %  \end{equation}
%and using the definition of $C$
%    \begin{equation}
%   - \frac{1}{2} (1-\eta )r_c^2 - 2c(R-r_c)  + (1 - \eta) L^2/4 = 0 \; .
% \end{equation}
% Dividing by $(1-\eta)/2$
%     \begin{equation}
%   - r_c^2 -  2 r_c (R-r_c)  + L^2/2 = 0 \; .
% \end{equation}
% thus
%  \begin{equation}
%   r_c^2 - 2 r_c R  + L^2/2 = 0 \; ,
% \end{equation}
% from which

% \begin{equation}
%   r^\dag = R - \sqrt{R^2 - L^2/2}  \; .
%   \label{eq:1dc}
% \end{equation}
 %where we assumed $r_c \leq L \leq R$. 
 
 For $r_c \geq r^\dag$ the optimal configuration is the one obtained for $\alpha^* =1/2$, whereas for $r_c \leq r^\dag$ the optimal configuration is the one corresponding to $\alpha^*=0$.
  
  In case (ii), we can repeat a similar derivation. We find that the maximum of the objective function is reached in  $\alpha = 1 - r_c/L$, and the function displays its minimum value either in $\alpha = 0$ or $\alpha = 1/2$. Also, here we find the critical value of Eq.~(\ref{eq:1d_meth}) where the optimal configuration of the fast layer changes from being perfectly symmetric to being asymmetric. Alternatively, we can determine the critical 
  length $L^\dag$ of the fast layer as shown in Eq.~(\ref{eq:1d}).

\begin{acknowledgements}
This project was partially supported by the Army Research Office under contract number W911NF-21-1-0194, by the Air Force Office of Scientific Research under award numbers FA9550-19-1-0391 and FA9550-21-1-0446, and by the National Science Foundation under award number 1927418.
The funders had no role in study design, data collection, and analysis, the decision to publish, or any opinions, findings, conclusions, or recommendations expressed in the manuscript.
\end{acknowledgements}

%\bibliography{references.bib} 


%%%%%%%%%%%%%%%

\newpage

\renewcommand{\thesection}{S\arabic{section}}
\setcounter{section}{0}
\renewcommand{\theequation}{S\arabic{equation}}
\setcounter{equation}{0}
\renewcommand{\thefigure}{S\arabic{figure}}
\setcounter{figure}{0}
\renewcommand{\thetable}{S\arabic{table}}
\setcounter{table}{0}

\section*{Supplementary Material}

\section{Multiplex transportation model}

\begin{figure}[!ht]
    \centering
    \includegraphics[width=0.3\textwidth]{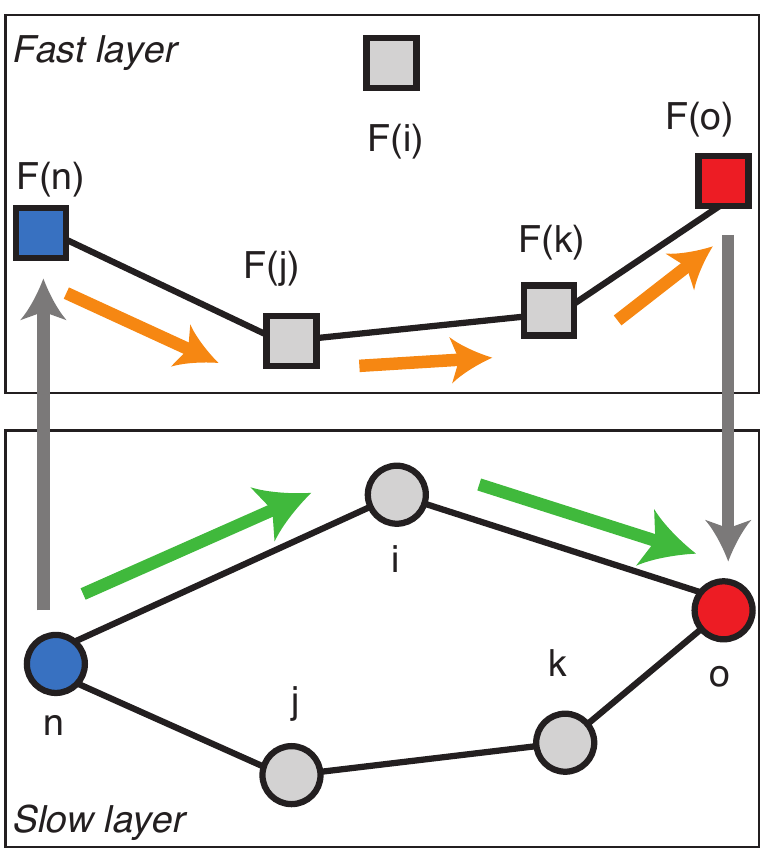}
    \caption{\textbf{Illustration of the multiplex transportation model.} 
    In the slow layer, the 
    %length 
    cost
    of an edge equals one;
in the fast layer, the 
%length 
cost
of an edge is reduced by the factor $0 \leq \eta
\leq 1$. Replica nodes across layers are connected by edges
with 
%length
cost
$c \geq 0$. Two possible paths connecting
node $n$ (blue circle) to node $o$ (red circle) are highlighted. 
The path $n \to i
\to o$ has 
%length
cost
equal to $2$ as it
uses only two edges in the slow layer (green arrows).
The second path, i.e., $n \to F(n) \to F(j) \to F(k) \to F(o) \to o$,
involves two changes of layer (gray arrows), and a path
in the fast layer (orange arrows). Its total 
%length 
cost
is $3 \eta + 2
c$, as each of the three edges used in the fast layer has 
%length
cost equal to $\eta$,
and $2c$ is the cost required to switch layer twice.
The second path is 
%shorter 
more convenient
than the first
one as long as $3 \eta + 2
c < 2$.}
    \label{fig:S0}
\end{figure}

\begin{table*}[!htb]
    \centering
    \begin{tabular}{r | r}
    notation &  object represented\\
    \hline
    \hline
         $\mathcal{G}$ &  set of nodes in the slow layer\\
         \hline
         $\mathcal{S}$ &  set of edges in the slow layer\\ 
         \hline
         $\mathcal{H}$ &  set of nodes in the fast layer\\
         \hline
         $\mathcal{F}$ &  set of edges in the fast layer\\ 
         \hline
         $N = |\mathcal{G}| = |\mathcal{H}|$ &  size of the network\\ 
         \hline
         for all $n \in \mathcal{G} \Leftrightarrow F(n) \in \mathcal{H}$ &  one-to-one map of nodes across layers\\ 
         \hline
         $w_{n,m} = 1$ for all $(n,m) \in \mathcal{S}$ &  
         %length 
         weight
         of the edges in the slow layer\\ 
         \hline
         $w_{F(n),F(m)} = \eta$ for all $(F(n),F(m)) \in \mathcal{F}$ &  %length
         weight
         of the edges in the fast layer\\ 
         \hline
         $w_{n,F(n)} = c$ for all $n \in \mathcal{G}$ & weight of interlayer edges or switching cost\\ 
         \hline
         $r_c = 2c/(1 - \eta)$  &  critical 
         %length 
         cost of the model, see Eq.~(\ref{eq:rs})\\ 
         \hline
          $d_{n \to m}$ for all $n,m \in \mathcal{G} \cup \mathcal{H}$  &  
          %length of the shortest path 
          cost of the minimum-cost
          from node $n$ to node $m$\\ 
         \hline
         $o \in \mathcal{G}$ &  center of the network\\ 
         \hline
          $d_n := d_{n \to o}$ for all $n \in \mathcal{G} \cup \mathcal{H}$ &  
          %length of the shortest path 
          cost of the minimum-cost path
          of node $n$ to $o$\\ 
         \hline
         $p_n$ for all $n \in \mathcal{G}$  &  weight of nodes in the slow layer\\ 
         \hline
         $\tau(\mathcal{F})$  & objective function, see Eq.~(\ref{eq:av_time})
         \\
         \hline
         $s_{n,m}(\mathcal{F})$ for all $(n,m) \in \mathcal{S}$ and $(F(n), F(m)) \notin \mathcal{F}$ & marginal gain in the objective function, see Eq.~(\ref{eq:gain})
         \\
         \hline
         $\mathcal{F}^*$ & optimal configuration of the fast layer, see Eq.~(\ref{eq:op})
         \\
         \hline
         $k^*$ & number of main branches of the optimal layer, see Eq.~(\ref{eq:geo_param})
         \\
         \hline\hline
         %\vdots & \vdots
    \end{tabular}
    \caption{List of variables and metrics used in the description of the multiplex transportation model and its associated optimization problem. For each of them, we report the notation used thorough the document and the quantity that is meant to represent.}
    \label{tab:defintions}
\end{table*}

We consider a multiplex network 
composed of a slow layer and a fast layer. 
For a schematic example of such a system, see Fig.~\ref{fig:S0}; for a list of variables/metrics used to characterize the model, see Table~\ref{tab:defintions}.
We denote with $\mathcal{G}$ the set of nodes in the slow layer, and with $\mathcal{H}$ the set of nodes in the fast layer. Both layers contain $N$ nodes, i.e., $|\mathcal{G}| = |\mathcal{H}| = N$. Each node in the slow layer has a one-to-one correspondence with a node in the fast layer; we indicate with $F(\cdot)$ the map between labels of nodes across layers, i.e., if $n \in \mathcal{G}$ then $F(n) \in \mathcal{H}$, and vice versa.
 We denote with $\mathcal{S}$ and $\mathcal{F}$ the set of edges of the slow and the fast layer, respectively. 
 %We have that $\mathcal{F} \subseteq \mathcal{S}$.
 Each edge in the fast layer has a replica edge in the slow layer, i.e., $(F(n), F(m)) \in \mathcal{F}$ implies $(n,m) \in \mathcal{S}$, but the vice versa is not necessarily true.
The weight of each edge in the slow layer equals one, i.e., $w_{n,m}=1$ for all $(n,m) \in \mathcal{S}$, whereas the weight associated with the edges of the fast layer is equal to $0 \leq \eta \leq 1$, i.e., $w_{F(n), F(m)} =\eta$ for all $(F(n),F(m)) \in \mathcal{F}$.  Replica nodes are connected to each other by  edges of weight $c \geq 0$, i.e., $w_{n,F(n)} = c$ for all $n \in \mathcal{G}$.

When considered in isolation, the slow layer form a single connected component, whereas the fast layer is not necessarily connected. The connectedness of the slow layer implies, however, that in the overall system, composed of the interconnected slow and fast layers, there exists at least a path connecting any pair of nodes.  The 
%length
cost
of a path, e.g., $n \to F(n) \to \ldots \to F(k) \to k \to \ldots \to m$, between two nodes $n$ and $m$ in the network is given by the sum of the 
%lengths
weights
of all edges that compose the path.  We denote with $d_{n \to m}$ the %length of the shortest 
cost of the minimum-cost
path between nodes $n$ and $m$. Naturally, the same definition of %length 
cost
applies to paths between any pairs of nodes, belonging to either the slow or the fast layer. For example, $d_{F(n) \to m}$ denotes the 
%length of the shortest 
cost of the minimum-cost
path between $F(n) \in \mathcal{H}$ and $m \in \mathcal{G}$. 
The 
%shortest 
minimum-cost
path between two nodes can either use edges in the slow layer only, or take advantage of some of the edges in the fast layer (see Figure~\ref{fig:S0}). In particular, the path $n \to F(n) \to F(m) \to \ldots \to F(r) \to F(s) \to s$ composed of $\ell$ edges in the fast layer only is preferred to its replica path $n \to m \to \ldots \to r \to s$ whenever $\ell$ is larger than the critical 
%length scale of the multiplex model, i.e.,
value
\begin{equation}
r_c = \frac{2 c}{ 1 - \eta} \; .
    \label{eq:Srs}
\end{equation}

\section{Optimization of the fast layer}

We identify a special node in the slow layer of the network, i.e., the center of the network.
The label of the center is $o$. We denote with $d_n := d_{n \to o}$ the 
%length of the shortest 
cost of the minimum-cost
path of the generic node $n$ to $o$. Also, we assume that each node $n$ in the slow layer has associated 
a weight $p_n \geq 0$, representing the demand of node $n$. 
We then define the weighted average 
%distance 
cost
to the center as
\begin{equation}
    \tau(\mathcal{F}) = \frac{\sum_{n \in \mathcal{G}} \, d_n \, p_n}{\sum_{n \in \mathcal{G}} \, p_n}  \; .
    \label{eq:Sav_time}
\end{equation}

We stress that the above function is computed over all nodes in the slow layer only, but eventual 
%shortest 
minimum-cost
paths can take advantage of edges in the fast layer. Clearly, $\tau$ depends on the various parameters of the model. In Eq.~(\ref{eq:Sav_time}), we explicit, on purpose, only the dependence of $\tau$ on the fast layer $\mathcal{F}$ as this is the primary object of our investigation. We consider in fact the optimization problem aimed at finding the best set of edges in the fast layer able to minimize the objective function of Eq.~(\ref{eq:Sav_time}). The minimization is constrained by the number of edges $L$ that are in the fast layer, where $L$ is measured in the same units of cost as $\tau$ and $c$. Specifically, we aim at solving

\begin{equation}
    \mathcal{F}^* = \arg \min_{|\mathcal{F}|=L} \, \tau(\mathcal{F}) \;, 
    \label{eq:Sop}
\end{equation}

where we indicated with $|\mathcal{F}|$ the number of edges in $\mathcal{F}$, and  
once more we did not write the explicit dependence of the 
objective function from the structure of the slow layer $\mathcal{S}$, the center of the network $o$, the weights $p_n$ for all nodes $n \in \mathcal{G}$, and the values of the parameters $\eta$ and $c$.

Finding the exact solution to the optimization problem of Eq.~(\ref{eq:Sop}) is computationally infeasible as it requires to test all possible ${|\mathcal{S}| \choose L}$ configurations of the fast layer. However, the number of suitable configurations for $\mathcal{F}^*$ can be restricted by some properties that the optimal layer must satisfy.

\section{Properties of the optimal fast layer}

We show that the optimal fast layer $\mathcal{F}^*$, solution of the optimization problem of Eq.~(\ref{eq:Sop}), is a tree containing at least one edge incident to node $F(o)$, i.e., the replica node of the center of the network.

\subsection*{Connectedness}

Assume that the edges in the fast layer $\mathcal{F}$ generate a graph consisting of at least one component that has no edges incident to node $F(o)$. Focus on one of these components, and 
indicate the set of its edges as $\mathcal{C} \subseteq \mathcal{F}$.

Consider now the edge 
$(F(m), F(n))  = \arg \max_{(F(i), F(j)) \in \mathcal{C}}  \max \{d_{F(i)}, d_{F(j)} \}$, 
i.e., the edge in the component that 
corresponds to the largest value of the cost of the minimum-cost path
%is at maximum distance from 
to the center $o$. 
Without loss of generality, suppose that $d_{F(m)} \geq d_{F(n)}$.
Based on our premise, the 
minimum-cost
%shortest 
path from node $F(m)$ to $o$ can be written as 
$F(m) \to F(n) \to \cdots \to F(s) \to F(r) \to r \to q \to \cdots \to o$, meaning 
there is at least one edge $(r,q) \in \mathcal{S}$ in the 
minimum-cost
%shortest 
path to $o$
such that $(F(r),F(q)) \notin \mathcal{F}$. 

Define the set $\mathcal{F}' = \mathcal{F} \setminus (F(m), F(n)) \cup (F(r), F(q))$, i.e., the same set as $\mathcal{F}$ but with the edge $(F(m), F(n))$ replaced by $(F(r), F(q))$. We have $\tau(\mathcal{F}') \leq \tau(\mathcal{F})$. In fact, the 
%length of the shortest 
cost of the minimum-cost
path to $o$ of every node $n \in \mathcal{S}$ whose 
%shortest 
minimum-cost
path to $o$ utilizes the edge $(F(m), F(n))$ when the fast layer is $\mathcal{F}$ will no increase when the fast layer is $\mathcal{F}'$; however, some of the other nodes that do not use the edge $(F(m), F(n))$ to reach $o$ when the fast layer is $\mathcal{F}$ can decrease %their shortest-path length 
the cost of their minimum-cost path
by using the edge $(F(r), F(q))$ when the fast layer is $\mathcal{F}'$.

We note that, when passing from $\mathcal{F}$ to $\mathcal{F}'$, the deletion of the edge $(F(m), F(n))$ does lead to any split of the component $\mathcal{C}$, except for the potential removal of node $F(m)$; however, the addition of the edge $(F(r), F(q))$ can lead to the merger of $\mathcal{C}$ with another component, as well as to the inclusion of an edge incident to $F(o)$. 

In summary, if the fast layer $\mathcal{F}$ is formed by at least one component that does not contain any edge incident to $F(o)$, then we can always find another configuration of the fast layer that is better than $\mathcal{F}$. The procedure can be iterated until its premise is no longer true. As a result, the optimal fast layer $\mathcal{F}^*$ must contain only one component with at least one edge incident to $F(o)$.

\subsection*{Tree structure}

Suppose the fast layer $\mathcal{F}$ contains a loop. 
If the loop is formed by an odd number of edges, then there is an edge $(F(n), F(m))$ for which $d_{F(n)} = d_{F(m)}$. This edge is irrelevant for $\tau(\mathcal{F})$,
%$d_i=d_j$. 
as no 
%shortest 
minimum-cost
path passes through it.  In fact, suppose node $i$ is such that $d_{i \to F(n)} \leq d_{i \to F(m)}$. 
%(here, $d_{r \to s}$ is the length of the shortest path from node $r$ to node $s$). 
Then, we can write $d_{i}  = d_{i \to F(n)} + d_{F(n)} \leq  d_{i \to F(m)} + d_{F(m) \to F(n)} + d_{F(n)}$. 
Similarly, if the loop has an even number of edges then there exists a node $F(n)$ such that there are two 
%shortest 
minimum-cost
paths from the node $F(n)$ to $o$. One of the edges incident to $F(n)$ in the loop can be removed without affecting $\tau(\mathcal{F})$. 
In both cases, the removed edge can be replaced by another edge  
potentially able to decrease $\tau$, thus the optimal fast layer $\mathcal{F}^*$ should not contain any loops.

\section{Optimization techniques}

In the previous sections, we proved that the 
optimization problem of Eq.~(\ref{eq:Sop}) can be solved by looking only at fast-layer configurations
consisting of trees that contain at least one edge incident to the replica of the center of the network.
This fact dramatically reduces the number of potential configurations that one should look at, however, it does not address the computational unfeasibility of the optimization problem. In this section, we introduce numerical techniques able to approximate solutions to the problem in an efficient and effective manner.

\subsection{Greedy optimization}

The algorithm takes as inputs the slow layer $\mathcal{S}$, the parameters $c$ and $\eta$, and the desired size $L$ of the fast layer. The output is the fast layer $\mathcal{F}_g$, representing a greedy solution to the optimization problem of Eq.~(\ref{eq:Sop}). 

We initially set $\mathcal{F}_g = \emptyset$, and we compute the %length of the shortest 
cost of the minimum-cost
path of all nodes to the center $o$. This information is stored in the vector $\vec{d}$. 
Also, we initialize $s_{n,m}(\mathcal{F}_g)$ for all edges $(n,m) \in \mathcal{S}$ such that $(F(n), F(m)) \notin \mathcal{F}_g$. $s_{n,m}(\mathcal{F}_g)$ quantifies the change in the objective function of Eq.~(\ref{eq:av_time}) that would be induced by adding the edge $(F(n), F(m)$ to $\mathcal{F}_g$, i.e., 
\begin{equation}
    s_{n,m}(\mathcal{F}_g) =  \tau(\mathcal{F}_g) - \tau(\mathcal{F}_g \cup (F(n),F(m))) \; .
    \label{eq:gain}
\end{equation}

Then, we iterate the following:

\begin{enumerate}

\item We solve
\begin{equation}
    (i, j) = \arg \max_{(n,m) \in \mathcal{S} | (F(n),F(m)) \notin \mathcal{F}_g} \, s_{n,m}(\mathcal{F}_g) \;,
    \label{eq:greedy}
\end{equation}
i.e., we find the edge corresponding to the largest 
drop in the objective function. Eventual ties are randomly broken.

\item We update $\mathcal{F}_g \to \mathcal{F}_g \cup (F(i),F(j))$. Also, we update the entries of the vector $\vec{d}$ using a suitably modified Dijkstra's algorithm.

\item If $|\mathcal{F}_g| = L$, we exit from the algorithm.

\item We update the scores $s_{n,m}(\mathcal{F}_g)$ for all $(n,m) \in \mathcal{S} $ such that $(F(n),F(m)) \notin \mathcal{F}_g$, and we go back to point 1. Please note that the update of each score $s_{n,m}(\mathcal{F}_g)$ also relies on the suitably modified Dijkstra's algorithm of point 2.

\end{enumerate}

The algorithm outputs not just the greedy solution $\mathcal{F}_g$, but also the 
%length of the shortest path 
cost of the minimum-cost path
of all nodes to the center $\vec{d}$ as well as the value of the objective function $\tau(\mathcal{F}_g)$.

\subsubsection*{Speeding up the greedy algorithm}

In the naive implementation described above, a significant number of computations are performed to update the scores $s_{n,m}(\mathcal{F}_g)$ for all edges in the slow layer. However, many of these updates are not required, and the algorithm can be speeded up quite significantly.

First, we know that the optimal fast layer is a tree containing at least one edge incident to $F(o)$, thus not all edges of the slow layer $\mathcal{S}$ should be considered at each stage of the algorithm. At the first iteration, only replica of edges that are incident to $F(o)$
should be considered; then in the following iterations, only replica edges that are incident to previously added edges or node 
$F(o)$ and that do not close eventual loops in the fast layer
should be considered.

Second and more important, there is no need to update the score of all potential edges that can be added. This follows from the fact that the score $s_{n,m}(\mathcal{F}_g)$ associated to the edge $(F(n),F(m)) \notin \mathcal{F}_g$, that is incident to another edge that has been already added to $\mathcal{F}_g$, does not increase as the number of iterations of the greedy algorithm increases. We therefore rely on a lazy-search procedure~\cite{minoux2005accelerated}. Specifically, we keep a sorted list (i.e., heap data structure) of the scores of all edges that are incident to edges in $\mathcal{F}_g$. When the new edge $(F(n),F(m))$ enters in the list, we compute its score $s_{n,m}(\mathcal{F}_g)$ and insert it in a temporary buffer. We then set $s_{\max} = s_{n,m}(\mathcal{F}_g)$. We pop out one edge at a time from the sorted list in descending order. Given an edge $(r,s)$, we update its score only if the current score $s_{r,q}(\mathcal{F}_g) > s_{\max}$. If updated, we insert the new score in the temporary buffer. If the updated score is $s_{r,q}(\mathcal{F}_g) \geq s_{\max}$, then we set $s_{\max} = s_{r,q}(\mathcal{F}_g)$ and go to the next element in the sorted list. Otherwise, we stop with the update, as we already found the edge with the largest score, i.e, the solution of Eq.~(\ref{eq:greedy}). We finally put back elements from the temporary buffer to the sorted list.

The only exception to the above rule 
of a score decreasing as the number of iterations of the algorithm increases is
for an edge that can potentially close a loop. However, we know that the optimal fast layer should not contain loops, thus edges of this type are not considered in our search.
%$s_{i,j} \leq s'_{i,j}$ is when $(i,j)$ is the only unperturbed edge in a loop of perturbed edges. 
%This fact can happen only if the perturbed edge is incident to $(i,j)$. Thus, the score of the edge $(i,j)$ is updated anyway according to the algorithm described above. We can avoid the insertion of edges that close loops.

If the score of $K$ edges $\{(n_1,m_1), \ldots, (n_K,m_K)\}$ is updated, we first compute their scores and insert them in the temporary buffer. Then, we find the one with the maximum score, say $(n,m)$. We simply set $s_{\max} = s_{n,m}(\mathcal{F}_g)$, and repeat the same instructions as above to perform our lazy search.

\subsubsection*{Modified Dijkstra's algorithm}

At step 2 of the greedy optimization algorithm, we mentioned that the
components of the vector of the 
%shortest-path lengths 
minimum-cost values
$\vec{d}$ are updated using a 
modified Dijkstra's algorithm. This procedure is used also to 
estimate potential changes to the vector $\vec{d}$, thus in the estimate of the 
scores $s_{n,m}(\mathcal{F}_g)$ at step 4 of the greedy optimization.
Our modified Dijkstra's algorithm is designed to perform local updates, as not all components of the vector $\vec{d}$ will necessarily change after the new edge $(F(i), F(j))$ is added to $\mathcal{F}_g$. Without loss of generality, let's assume that before the addition of the edge, node $F(i)$ has degree equal zero in the fast layer, whereas the degree of node $F(j)$ is larger than zero.
We first update  $d_{F(i)} \to  \min \{ d_{F(i)}, d_i+c, \min_{(i,q) \in \mathcal{F}_g} d_{F(q)} + \eta \}$. The three elements corresponds respectively to: (i) unchanged 
cost of the minimum-cost
%length of the shortest 
path,
(ii) 
%shortest 
cost of the minimum-cost
path reduced by taking the slow layer, (iii) 
%shortest 
cost of the minimum-cost
path reduced by taking the fast layer. We then update the components of the other nodes using a Dijkstra-like algorithm. The algorithm is started from node $F(i)$, and exploits edges in both $\mathcal{S}$ and $\mathcal{F}_g$ leading therefore to updates in the vector $\vec{d}$ that regards nodes in both the slow and the fast layer. However, it considers only moves $q \to p$ such that $d_q + w_{q,p} \leq d_p$; in such a case, the component of the node $p$ is updated as $d_p \to  d_q + w_{q,p}$.  The  Dijkstra-like algorithm does not necessarily visit all nodes in the network, but only those whose component in the vector $\vec{d}$ is decreased by the addition of the edge $(F(i),F(j))$.

\subsubsection*{Submodularity of the objective function}

If the fast layer is composed of a single connected component, 
the score $s_{n,m}(\mathcal{F}_g)$ associated with each potential edge $(F(n),F(m))$ 
that could be added to the set $\mathcal{F}_g$ is a non-increasing function of the number of edges already added to $\mathcal{F}_g$.  
This fact follows from the simple observation that the only effect that the addition of the edge $(F(n),F(m))$ can have is adding novel %shortest 
minimum-cost
paths in the network, potentially reducing the 
%length of the shortest 
cost of the minimum-cost
path of some nodes. The drop in the 
%shortest-path length 
cost of the minimum-cost path
induced by the addition of the edge $(F(n),F(m))$ is maximal when $\mathcal{F}_g = \emptyset$. If $\mathcal{F}_g \neq \emptyset$, however, the drop is potentially reduced given that the other edges already in the set can provide 
%shortest 
minimum-cost
paths to the center that do not pass through the edge $(F(n),F(m))$.

As a matter of fact, we can write

\begin{equation}
\begin{array}{ll}
  s_{n,m} (\mathcal{F}') & =     \tau(\mathcal{F}')  - \tau(\mathcal{F}' \cup (F(n),F(m))) 
  \\ & \geq  \tau(\mathcal{F}'')  - \tau(\mathcal{F}'' \cup (F(n),F(m)))
  \\ & = s_{n,m} (\mathcal{F}'') 
  \end{array}
  \; ,
  \label{eq:submodularity}
\end{equation}

where $\mathcal{F}' \subseteq \mathcal{F}''$, both $\mathcal{F}'$ and $\mathcal{F}''$ are trees with at least one edge incident to $F(o)$, and $(F(n),F(m))$ is an arbitrary edge that is incident to one edge in both $\mathcal{F}'$ and $\mathcal{F}''$. 

In summary, the function $\tau$ that is optimized is a non-negative, non-increasing, submodular function. As such, the proposed greedy algorithm allows us to find solutions that are at maximum a factor $(1-1/e) \simeq 0.63$ above the ground-truth optimum~\cite{nemhauser1978analysis}.

The only exception to the inequality~(\ref{eq:submodularity}) is when both $F(n)$ and $F(m)$ are adjacent to edges in $\mathcal{F}''$. This is, however, not possible as the edge $(F(n), F(m))$ would close a loop, which is in contradiction with the fact that the optimal fast layer must be a tree.

\subsubsection*{Selecting greedy solutions}
%\subsubsection*{Greedy solutions for $r_c \geq 1$}

In the above formulation of the greedy optimization algorithm, we tacitly assumed
that $r_c < 1$, meaning that the addition of every edge to the fast layer can potentially reduce the value of the objective function. This fact allows us to initialize the algorithm with $\mathcal{F}_g = \emptyset$.

A simple way to re-use the previous algorithm for $r_c \geq 1$ is changing the initialization. Specifically, we can proceed by first identifying a node $n \in \mathcal{G}$ for which $d_n = \lfloor r_c \rfloor$ and one of the 
%shortest 
minimum-cost
paths connecting $n$ to $o$, and then adding $(F(i),F(j))$ to $\mathcal{F}_g$ for each edge $(i,j) \in \mathcal{S}$ that is part of such a 
%shortest
minimum-cost
path.

We adopt a different protocol that can be used for any value of $r_c$. Although the switching cost $c$ is an input of the optimization problem, we treat it as a variable by considering $M = 1,000$ different values in the interval $[0, (1-\eta)/2]$. For each of them, we find a solution using the greedy optimization algorithm, namely $\mathcal{F}_g^{(1)}, \ldots, \mathcal{F}_g^{(M)}$.
We compute the value of the objective function associated to each of these sets
by using the input value of the switching cost $c$, and 
find the best solution by identifying 
the one corresponding to the smallest value of the objective function. 

\subsubsection*{Complexity of the greedy algorithm}

\begin{figure}[!htb]
\includegraphics[width=0.45\textwidth]{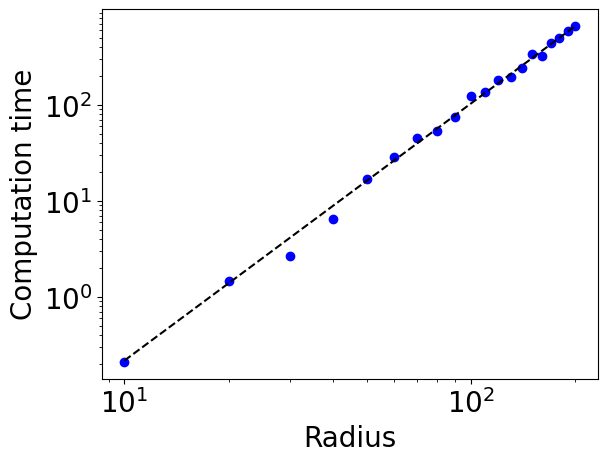}
\caption{\textbf{Computational complexity of the greedy algorithm.} We generate slow layers according to our multiplex transportation model with variable 
 triangular lattice radius $R$. We set the parameters of the model as $L=R$, $\eta=0.1$, and $c=0.1$. We then apply the greedy algorithm to approximates solutions to the optimization problem of Eq.~(\ref{eq:Sop}). We plot the computational time $V$ required by the greedy algorithm as a function of $R$. Time is measured in seconds. Simulations were run
on an Intel(R) Xeon(R) CPU E5-2690 v4 @ 2.60 GHz. 
The dashed line stands for $V \sim R^{2.7}$.
%denotes the fit for $4.59 \times 10^{-4} R^{2.7}$.
}
\label{fig:tc}
\end{figure}

We show the computational complexity of the greedy algorithm after implementing the speed-ups described above. We use the two-dimensional multiplex transportation model introduced in the main manuscript with the radius of the triangular lattice equal to $R$. We generate greedy solutions for $c=0.1$, $\eta=0.1$, and $L=R$. The time required to obtain the greedy solution is plotted against the radius $R$ in Figure~\ref{fig:tc}. Note that the multiplex transportation model with radius $R$ has $N = 3R^2 +3R +1$ nodes in each layer. 

\subsection{Simulated Annealing}

We use a simulated-annealing scheme to approximate solutions to the optimization problem of Eq.~(\ref{eq:Sop}). Specifically, we start from a set $\mathcal{F}_{\textrm{sa}}$ constructed by sequentially adding $L$ randomly chosen edges to the fast layer such that no loops are present, the graph is connected and at least one edge is incident to node $F(o)$. We estimate the objective function $\tau(\mathcal{F}_{\textrm{sa}})$. We impose the cooling factor $\gamma = 0.999$, the minimum temperature $T_{\min} = 10^{-3}$, and the initial temperature $T = 10^{2}$. We then iterate:

\begin{enumerate}
    \item Select a random edge $(F(n), F(m)) \in \mathcal{F}_{\textrm{sa}}$ such that either the degree of node $F(n)$ or $F(m)$ equals one in the fast layer. Select a random edge $(i,j) \in \mathcal{S}$ such that the sum of the degrees of nodes $F(i)$ and $F(j)$ in the fast layer is equal to one.

    \item Consider the set $\mathcal{F}'_{\textrm{sa}} = \mathcal{F}_{\textrm{sa}} \setminus (F(n), F(m)) \cup (F(i), F(j))$. Please note that $\mathcal{F}'_{\textrm{sa}}$ is still a compatible solution of the optimization problem being a connected tree composed of $L$ edges with at least one edge incident to node $F(o)$. Compute $\tau(\mathcal{F}'_{\textrm{sa}})$.

    \item With probability $\min\{ 1, e^{[\tau(\mathcal{F}'_{\textrm{sa}}) - \tau( \mathcal{F}_{\textrm{sa}})]/T}  \}$, we accept the change and update $\mathcal{F}_{\textrm{sa}} \to \mathcal{F}'_{\textrm{sa}}$.

    \item Update the temperature as $T \to \gamma \, T$. If $T > T_{\min}$, go back to point 1, otherwise end the algorithm.
    
\end{enumerate}

\section{Results}

\begin{comment}
\subsection{Characterization of the optimal fast layer}

The vast majority of our analysis focuses on the case $|\mathcal{F}| \ll |\mathcal{S}|$, i.e., the number of edges $L$ of the optimal fast layer is much smaller than the number of edges in the slow layer. In this regime, the geometry of the optimal layer can be characterized by the single parameter
\begin{equation}
    k^* = \sum_{(F(n), F(m)) \in \mathcal{F}^*} \, \left[ \delta_{F(o), F(m)} + \delta_{F(n), F(o)} \right] \; ,
    \label{eq:geo_param}
\end{equation}
which corresponds to the total number of edges that are incident to the node $F(o)$. In Eq.~(\ref{eq:geo_param}), we use the Kronecker's delta function $\delta_{x,y} = 1$ if $x=y$ and $\delta_{x,y} =0$ otherwise.
$k^*$ quantifies the number of main branches that depart from the center of the system. Additional ramifications are not accounted for as they are not observed in the regime $|\mathcal{F}| \ll |\mathcal{S}|$. Based on our analytical derivations, we know that $k^* \geq 1$. We numerically study how its value depends on the geometry of the slow layer and the other parameters that characterize the multiplex transportation model.

\end{comment}

\subsection{Erd\H{o}s-R\'enyi graphs}

\begin{figure}[!htb]
\includegraphics[width=0.45\textwidth]{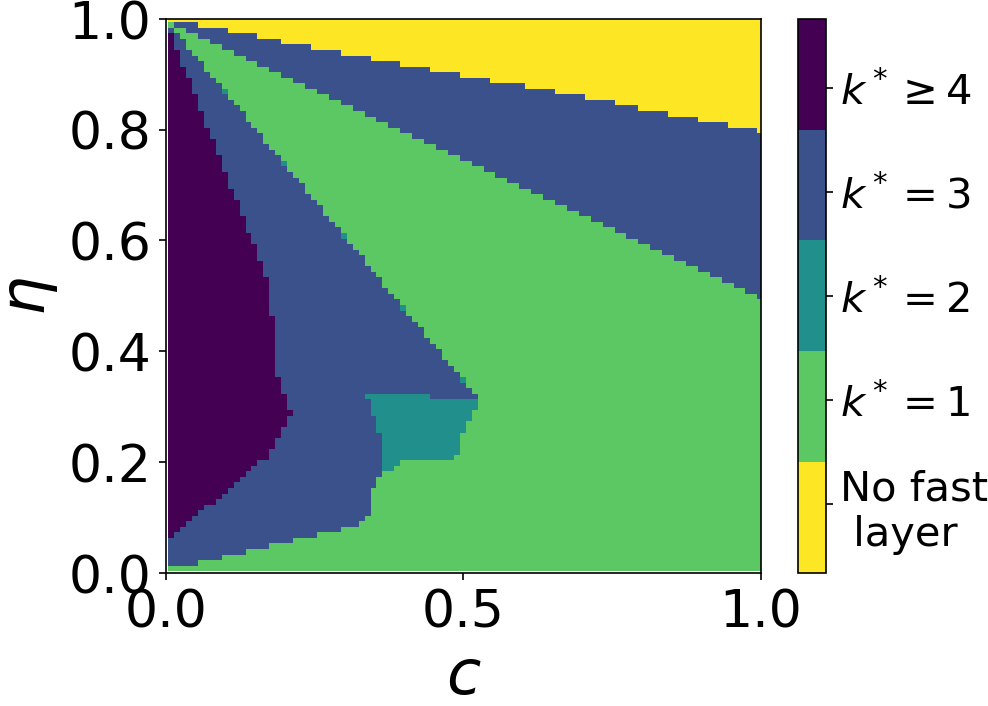}
\caption{\textbf{Phase diagram for random networks.} We show the heat map illustrating the type of optimal solutions obtained for different combinations of $c$ and $\eta$. We consider a slow layer consisting of an ER network of size $N=1000$ and set the 
%length 
size
of the fast layer solution to $L=10$. 
We plot the number of branches $k^*$ observed in the optimal fast layer as a function of the model parameters $\eta$ and $c$.
The yellow region indicates $L \leq r_c$. In this regime, 
%that it 
there is no benefit in building a fast layer.
}
\label{fig:er}
\end{figure}

In Figure~\ref{fig:er}, we consider a slow layer formed by the edges of a single instance of the Erd\H{o}s-R\'enyi (ER) model with $N=1, 000$ and average degree $\langle k \rangle=4$. As the center of the network, we use the node with the largest degree equal to $11$ in this specific realization of the ER model. In the objective function of Eq.~(\ref{eq:av_time}), we assume a weight $p_n = $const. for all nodes $n \in \mathcal{G}$. We use the greedy optimization algorithm to approximate solutions to the optimization problem of Eq.~(\ref{eq:Sop}) for $L=10$ and various combinations of the model parameters $c$ and $\eta$.  The heat map displays the corresponding values of $k^*$ as a function of the model parameters. No optimal fast layer can be constructed if $L \leq r_c$. For $L> r_c$, we observe instead a rich phase diagram where $k^*$ ranges between $1$ and $5$. In general, for a fixed value of $\eta$ and sufficiently large $c$, we always observe $k^*=1$.

\subsection{Two-dimensional lattices}
\label{sec:2dlattice}

\begin{comment}
The vast majority of the results reported in this paper are obtained for slow layers generated from triangular lattices. Specifically, one site of the lattice is selected as the center of the reference system; the coordinates of all other sites of the lattice are in the form $(a+\frac{b}{2}, \frac{\sqrt{3}b}{2})$ for integer values of $a$ and $b$ such that $|a|+|b|\leq R$, where $R$ is the radius of the triangular lattice. As a result, the boundaries of the lattice give the system a hexagonal shape. The sites of the lattice are the nodes of the slow layer; each pair of nodes in the slow layer is connected if the corresponding sites in the triangular lattice are at distance one in the lattice. 
\end{comment}

\begin{figure*}[!htb]
\includegraphics[width=0.85\textwidth]{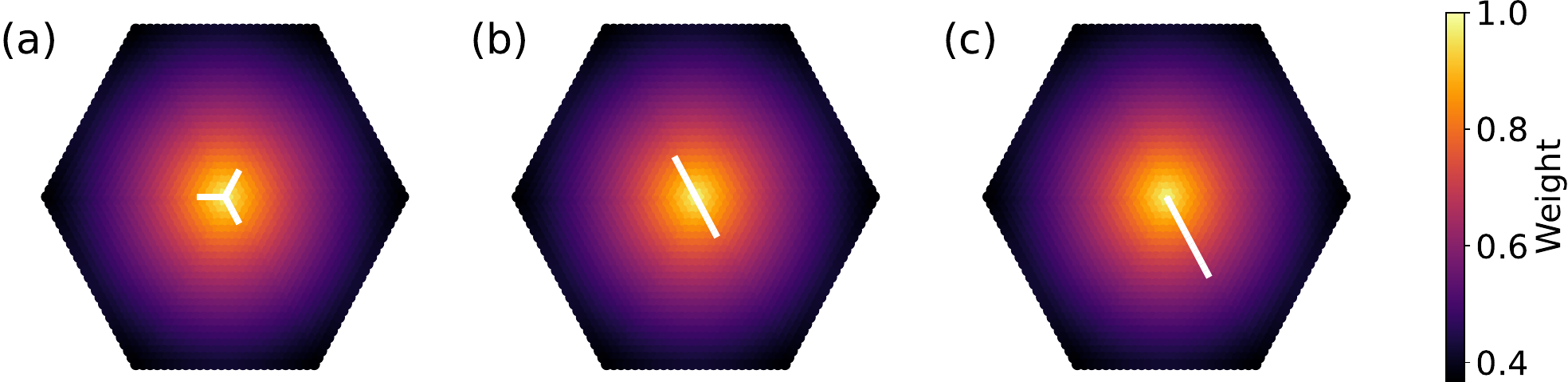}
\caption{\textbf{Two-dimensional triangular lattices with exponentially decaying weights.} (a) Optimal solution for the system with the fast layer composed of three branches. We use $R=25$, $c=0.1$, $\eta=0.1$, and $L=12$. The weight associated to the individual nodes in the slow layer is represented by the color map. (b) Same as in (a), but for $c=0.7$. (c) Same as in (b), but for $c=1.2$.}
  \label{fig:den1}
\end{figure*}

\begin{figure}[!htb]
\includegraphics[width=0.5\textwidth]{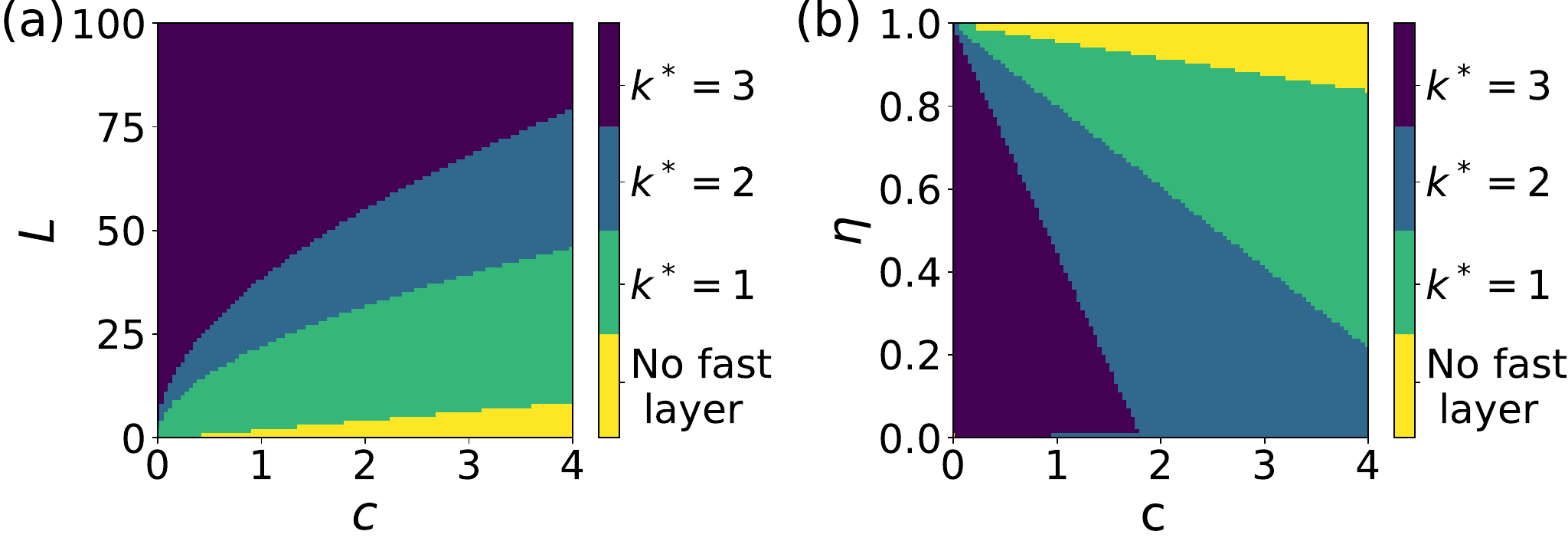}
\caption{\textbf{Phase diagrams for optimal fast-layer configurations in two-dimensional triangular lattices with exponentially decaying weights.} (a) We show the heat map illustrating the type of optimal solutions obtained for different combinations of $c$ and $L$. We consider a slow layer with $R=100$ and set the weight of the edges in the fast layer to be $\eta=0.1$. 
We plot $k^*$ as a function of $c$ and $L$.
The yellow region indicates that it is not beneficial to build a fast layer for the values of $c$ and $L$ as it will not reduce the 
%shortest
minimum-cost path to the center regardless of the structure, i.e., $L<r_c$. (b) same as in (e), but the y-axis is changed to $\eta$ and the 
%length 
size
of the fast layer is set to $L=50$.}
  \label{fig:den2}
\end{figure}

In Figure~\ref{fig:den1}, we report results concerning the optimal fast-layer configuration obtained on two-dimensional lattices when the weight $p_n$ associated to node $n \in \mathcal{G}$ is
\begin{equation}
p_n \sim e^{-d(n,o)} \;, 
\label{eq:expo}
\end{equation}
where $d(i,o)$ is the Euclidean distance between nodes $n$ and $o$.
We generate phase diagrams similar to Figures 2 e and f in the main paper but for weights obeying Eq.~(\ref{eq:expo}), see Figure~\ref{fig:den2}.

\section{Continuous-space approximations}

\subsection{Star-like systems}

\begin{figure}[!htb]
\includegraphics[width=0.45\textwidth]{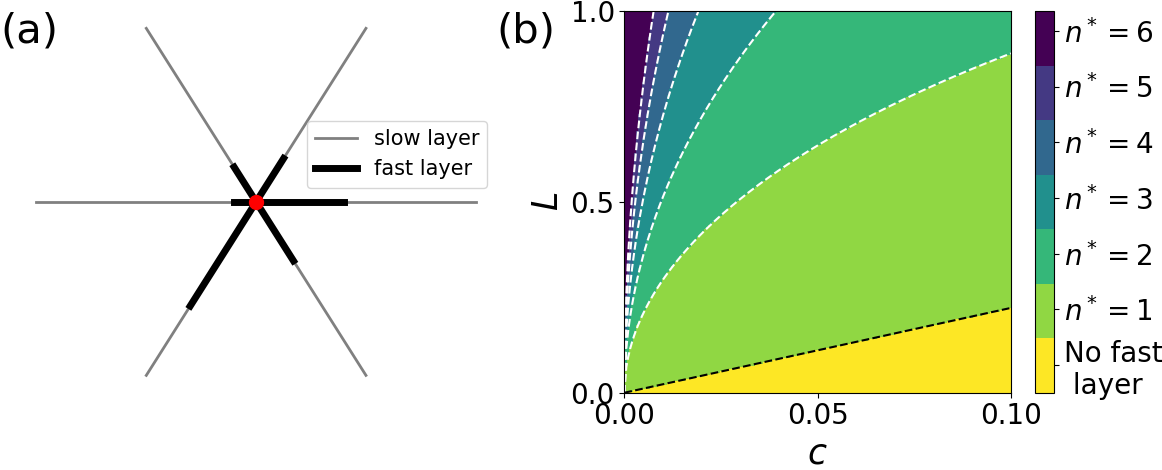}
\caption{\textbf{Star-like systems}. (a) Slow layer composed of $q=6$ branches with a fast layer composed of $n=6$
  branches of unequal length. The red circle shows the center of the system. (b) Optimal solutions for the system shown in (a) for different values of $L$ and $c$ and fixed $R=1$ and $\eta=0.1$. The dashed white curves show the analytically obtained boundaries between different solutions, i.e., Eq.~(\ref{eq:nb3}), and the black dashed line shows the region where no fast layer solution exists, i.e., $L \leq r_c$.}
  \label{fig:nbranches}
\end{figure}

We consider a slow layer consisting of $q$ branches of length $R$ emerging from the center $o$. The optimal fast layer must be a tree rooted in the replica of the center with $1 \leq n^*\leq q$ branches (see Figure \ref{fig:nbranches}). In this setting, we can show that the length of these $n^*$ branches must be equal. 
%First, it is easy to see that all the branches in the optimal solution must be at least of length $r_c$. Branches of length shorter than $r_c$ will not be used. 

Let us assume that this is not true and the optimal solution consists of branches of potentially unequal lengths. Consider two of them of length $\ell_1$ and $\ell_2$, respectively. We can map this configuration to the one-dimensional case by simply using $L=\ell_1 + \ell_2$. This automatically tells us that the configuration in which $\ell_1 = \ell_2$ is better than any configuration where $\ell_1 \neq \ell_2$.  We note in fact that the inequality $r_c \geq r^\dag$ is already satisfied, otherwise, the optimal solution would have consisted of a number of branches different from $n^*$.

The above consideration is valid for any pair selected out of the $n^*$ branches, therefore, the optimal solution in this setting should be given by branches of equal length.

We can obtain the boundaries between the optimal fast-layer solutions for this system by extending the framework used in the one-dimensional case. As shown above, the solutions can be characterized by the number of branches $n^*$ of equal length $L/n^*$. Similar to Eq.(7) of the main paper, the average cost to reach the center for a single branch with a fast layer of length $\ell$ attached to the center is given by

\begin{equation}
  \tau_{\text{single}}(\ell) =
  \left\{
    \begin{array}{ll}
      R^2/2 & \textrm{ if } 0 \leq \ell \leq r_c
              \\
      C + (1- \eta)  \ell \left[ \ell /2 -  R \right] & \textrm{oth.}
    \end{array}
    \right.
      \;, 
      \label{eq:nb1}    
    \end{equation}
    where
\begin{equation}
C = \frac{1}{2} (1-\eta )r_c^2 + 2c(R-r_c) +\frac{1}{2}R^2 \; .
\label{eq:nb2}    
\end{equation}

Therefore, the average cost to reach the center for a generic solution with $n^*$ branches can be written using Eq.~\ref{eq:nb1} as

 \begin{equation}
  \tau(L,n^*) =n^* \tau_\text{single}(L/n^*)+(q-n^*)\tau_\text{single}(0)
      \label{eq:nb3}    
    \end{equation}

Consequently, the boundary between solutions with $n^*$ and $n^*+1$ branches is given by the condition $\tau(L,n^*) = \tau(L,n^*+1)$. We obtain the following critical values 

 \begin{equation}
  L^\dag(n^*) =\sqrt{ n^*(n^*+1)(2 R r_c - r_c^2   ) } \; .
      \label{eq:nb4}    
    \end{equation}

The obtained boundaries alongside the optimal solutions for $\eta=0.1$, $R=1$, and different values of $c$ and $L$ are shown in Figure~\ref{fig:nbranches} b.

\subsection{Two-dimensional systems}

\begin{figure*}[!htb]
\includegraphics[width=0.85\textwidth]{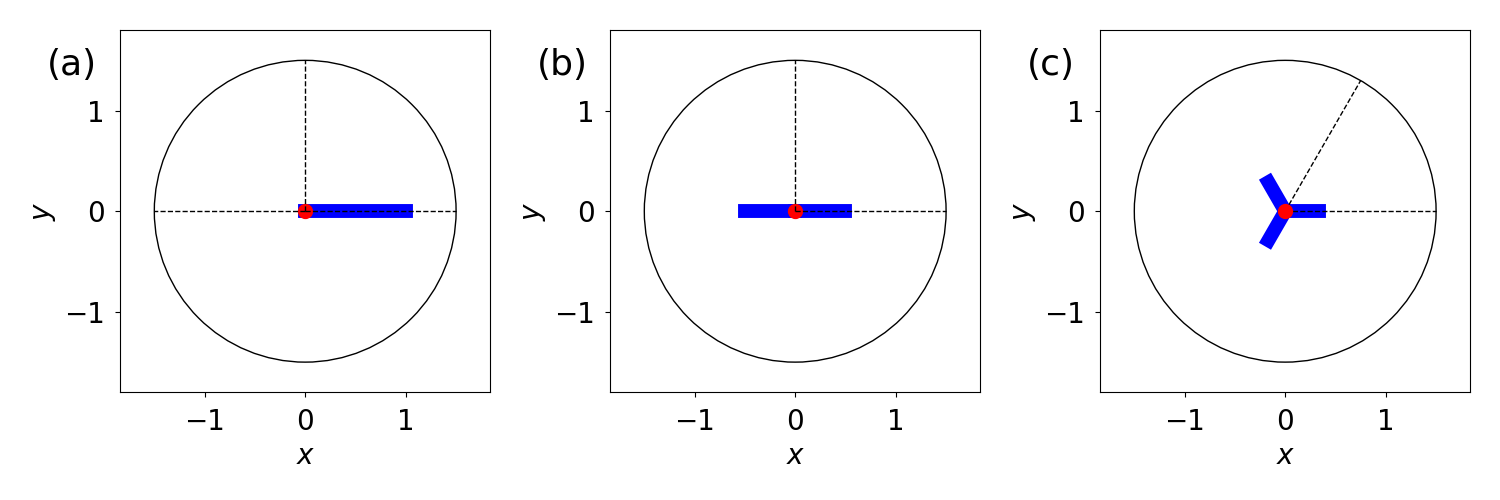}
\caption{(a) Circle with a fast layer composed of a single
  branch. Here, $R=1.5$ and $L=1$.
  The two relevant quadrants for the calculations of the
  objective function
  are
  delimited by the dashed lines. (b) Same as in (a), but for $n=2$
  branches. Only one quadrant is relevant for the computation of the
  objective function 
  and its delimited by the dashed lines.
  (c) Same as in (b), but for $n=3$ branches.}
  \label{fig:ill}
\end{figure*}

We consider a circle of radius $R$. A fast layer of linear size $L$
is present. 
We assume that the fast layer is composed of $n$ segments or
branches, each of length $L/n$. 
Each segment departs from the origin.
Consecutive segments are separated 
one from the other by an angle equal to $2\pi/n$ (see Figure~\ref{fig:ill}).
Please note the above assumption is reasonable, but
we do not have a mathematical proof that the optimal 
fast layer has such a geometry.
Our goal is to compute the 
%shortest 
cost of the minimum-cost
path $\tau(n)$
of all points in the slow layer to the center of the circle, and
then determine the number of branches of the optimal fast layer as
\begin{equation}
n^* = \arg \, \min_{n} \, \tau(n) \; .
  \label{eq:opt_branches}
\end{equation}

No fast layer is created if $R \leq r_c$. 
In such a case, the objective function reads
\begin{equation}
    \tau(n=0) = 4 \tau_{\textrm{e}} \; . 
    \label{eq:n0}
\end{equation}
If $R \geq r_c$, the solution should 
be determined by comparing the value of the objective function
for different $n$ values. These can be systematically computed as follows.
For $n=1$, we have
\begin{equation}
    \tau(n=1) = 2 \tau_{\textrm{e}}  + 2 \tau_{\textrm{f}} (L, \pi/2) \; , 
    \label{eq:n1}
\end{equation}
which stands for the sum of the 
%lengths of all shortest
costs of all minimum-cost
paths in two empty quarters of the circle, i.e., spanning an angle equal to
$\pi/2$, and of the 
%lengths of all shortest 
costs of all minimum-cost
paths of two quarters containing a fast layer of length $L$ located on one of the sides of the quarter
of the circle. 
For $n > 1$,  we have
\begin{equation}
    \tau(n) = 2n  \tau_{\textrm{f}} (L/n, \pi/n) \; , 
    \label{eq:n2}
\end{equation}
thus the objective function
is given by the sum of $2n$ identical contributions. Each
contribution refers to the 
%length of the shortest paths
cost of the minimum-cost paths
associated to a section of the circle that spans
an angle equal to $\pi/n$ and takes advantage of a fast layer of
length $L/n$ that is located on one of the sides of the section.

\begin{figure}[!htb]
\includegraphics[width=0.45\textwidth]{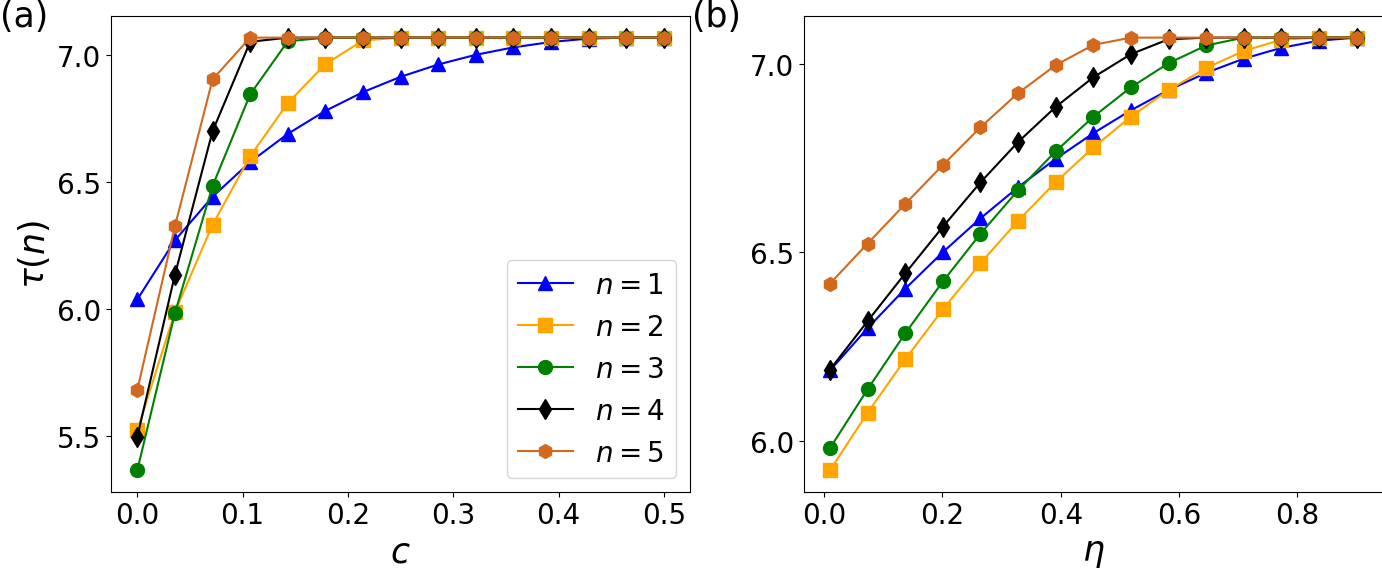}
\caption{(a) $R=1.5$, $L=1$, $\eta=0.1$.
  We plot $\tau(n)$ as a function of
  $c$ for different $n$ values. (b) Same as in (a)
   but for fixed $c = 0.05$ and varying $\eta$.}
  \label{fig:S2}
\end{figure}

\begin{figure}[!htb]
\includegraphics[width=0.45\textwidth]{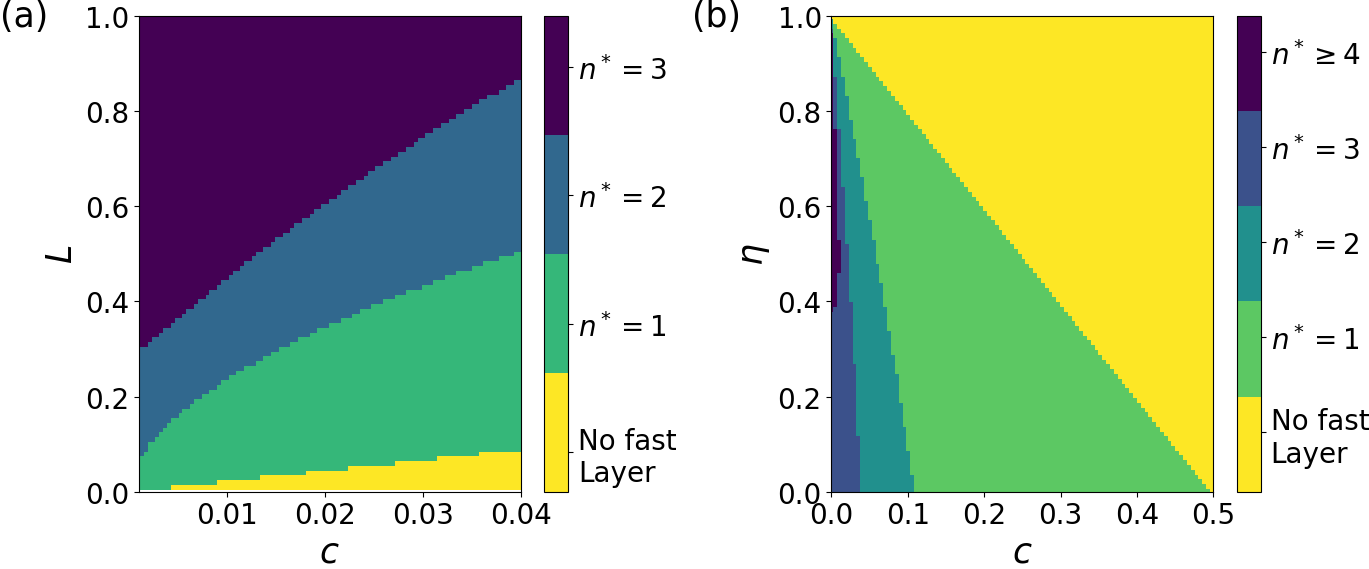}
\caption{(a) $R=1.5$ and $\eta=0.1$. For each pair of values
  $c$ and $\eta$, we determine the value of $n$ solution of Eq.~(\ref{eq:opt_branches}). (b) same as (a) but $L=1$ and $c$ and $\eta$ are varied. }
  \label{fig:S3}
\end{figure}

The quantities appearing on the r.h.s. of Eqs~(\ref{eq:n1}) and~(\ref{eq:n2}) can be computed by solving specific integrals. Details are reported in the following sections.
From Figure~\ref{fig:S2}(a), we see that as  long as $c$ increases and $\eta$ is kept constant, $n^*$ decreases.
We see instead from Figure~\ref{fig:S2}(b), that $n^*$ increases as $\eta$ increases for fixed $c$.
The result is consistent with what already observed in the one-dimensional and on ER graphs, and the heat maps of Figure~\ref{fig:S3} confirm the fact that qualitatively similar results are valid regardless of the dimensionality of the system.

In Figure~\ref{fig:3}, we plot $n^*$, as defined in
Eq.~\ref{eq:opt_branches}, as a function of $\eta$ and $c$. Only
values $1 \leq n \leq 6$ is considered in the numerical test.

\subsubsection*{Empty quarter}
This is the simplest case where the fast layer is not used.
We simply have
%Travel time is simply given by

\begin{equation}
  %\tilde{T}(R) = 
  \tau_{\textrm{e}} = 
  \int_0^R \, dr \; \int_0^{\pi/2} \, d\theta \;r =
  \frac{\pi \, R^2}{4}\; .
  \label{eq:empty_circle}
\end{equation}

%The integrand $t_{l}(r,\theta) = r$ is
%the Euclidean distance to the origin of the point $(r,\theta)$.

\subsubsection*{Section of the circle spanning an angle $\phi$ with a fast
  layer of length $\ell$}
The fast layer is a segment going
from $(0,0)$ to $(\ell, 0)$.
A generic point $(r,\theta)$, with $0 \leq r \leq R$ and $0 \leq \theta
\leq \phi$  can reach the origin either using or not the
fast layer.
If it does not use it, then 
the cost of the minimum-cost path
%the shortest-path length
is $q_l = r$. If it does use it, then the 
%shortest-path length
cost is
\[
  q_{g}  = \sqrt{r_g^2 + r^2  - 2
    r \, r_g \, \cos \theta\  }  + \eta r_g + 2 c \; .
  \]

The first term is the distance of $(r,\theta)$ to the point $(r_g,0)$ on
the fast layer that must be reached using the slow layer.
The other two terms account for the cost of the path 
to the center on the fast layer and the penalty associated with the
change of layers. We have that $r_g = \min \{ \ell, r_\times \}$, with
\begin{equation}
  r_\times = \arg \, \min_{z}  \,  \sqrt{z^2 + r^2  - 2
  r \, z \, \cos \theta\  }  + \eta z  \; .
  \label{eq:fast1_circle}
\end{equation}

If $r_g = \ell$, the minimum-cost path takes advantage of the entire
length of the fast layer; if $r_g = r_\times \leq \ell$, then only part
of the fast layer is used by the minimum-cost path to the center.
To compute $r_\times$, we just find the value for which
\[
  \frac{d}{dz} \left( \sqrt{z^2 + r^2  - 2
  r \, z \, \cos \theta\  }  + \eta z \right) = 0 \; .
\]

%For shortness of notation, let's use $r_x = r \cos \theta$, then
%we can write
%\[
%  \frac{d}{dz} \left( \sqrt{z^2 + r^2  - 2
%  r_x \, z }  + \eta z \right) = 0 \; .
%\]

%We get
%\[
%  \frac{r_\times - r_x} {\sqrt{ r_\times^2 + r^2  - 2
%  r_x \, r_\times  } }  + \eta = 0
%\]
%then
%\[
%  (r_\times - r_x)^2 = \eta^2 (r_\times^2 + r^2  - 2
%  r_x \, r_\times  )
%\]

%\[
%  (r_\times - r_x)^2  = \eta^2 (r_\times^2 + r^2  - 2
%  r_x \, r_\times  + r_x^2 - r_x^2 )
%\]

%\[
%  (r_\times - r_x)^2  = \eta^2 (r_\times^2 + r_x^2  - 2
%  r_x \, r_\times ) + \eta^2 (r^2 - r_x^2 )
%\]

%\[
%  (r_\times - r_x)^2  = \eta^2 (r_\times - r_x)^2 + \eta^2 (r^2 - r_x^2 )
%\]

%\[
%  (r_\times - r_x)^2 (1-\eta^2)  = \eta^2 (r^2 - r_x^2 )
%\]

%\[
%   (r_x - r_\times) \sqrt{1-\eta^2}  = \eta \sqrt{r^2 - r_x^2 } \; ,
% \]
% since we are interested in the case $r_\times \leq r_x \leq r$.
% We then can write
%\[
%  r_x - r_\times  =  \sqrt{ (r^2 - r_x^2) \, \eta^2/(1- \eta^2) }
%\]

%\[
%r_\times = r_x - \sqrt{ (r^2 - r_x^2) \,  \eta^2/(1- \eta^2) } \; .
%\]
%Going back to the original notation, we can write
%\[
%  r_\times = r \cos \theta - \sqrt{ r^2 (1- \cos^2 \theta) \, \eta^2/(1- \eta^2)    }
%\]
%\[
%  r_\times = r \cos \theta - \sqrt{ r^2 \sin^2 \theta \, \eta^2/(1- \eta^2)
%  } \; ,
%\]
%where we use the fact that $\cos^2 \theta + \sin^2 \theta=1$,
%and finally we can write
After some calculations, we arrive to the expression
\begin{equation}
r_\times = r \left(  \cos \theta - \frac{\eta}{\sqrt{1 - \eta^2}} \,  \sin \theta \right) \; .
  \label{min:r}
\end{equation}

We can now insert the expression of Eq.~(\ref{min:r}) into
Eq.~(\ref{eq:fast1_circle}) to determine
the value of $q_{g}$ in $r_\times$.

After some calculations, we find
\begin{equation}
  q_{g} = \left\{
    \begin{array}{ll}
q_{g_1} & \textrm{ if } \ell \leq r \left( \cos \theta - \frac{\eta}{\sqrt{1-\eta^2}}
          \sin \theta \right)
      \\
      q_{g_2} & \textrm{ otherwise }
    \end{array} \right.
    \label{eq:tglo}
    \end{equation}
    with
  \begin{equation}
q_{g_1} =  \sqrt{\ell^2 + r^2 - 2 r \, \ell \cos \theta} + \eta \ell +
2 c
\label{eq:tglo1}
      \end{equation}
and
\begin{equation}
  q_{g_2} =
    r \left( \eta  \cos \theta + \sqrt{1- \eta^2} \sin \theta
\right) + 2 c \; .
\label{eq:tglo2}
\end{equation}

 To compute our quantity of interest, we need to perform the integral
 \begin{equation}
\tau_{\textrm{f}} (\ell, \phi) = \int_0^R dr \, \int_0^\phi d\theta \; \min \{ r, q_{g} \} \; .
\label{eq:tglo_final}
\end{equation}

\begin{figure}[!htb]
\includegraphics[width=0.45\textwidth]{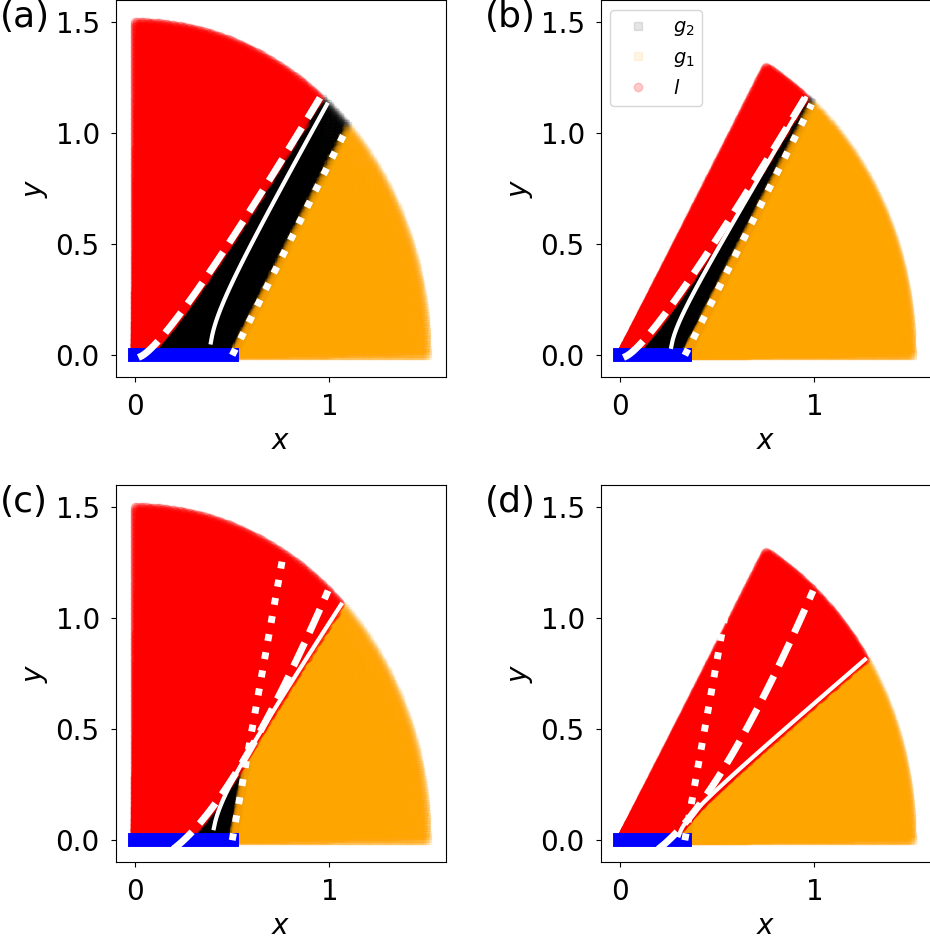}
\caption{(a) $R=1.5$, $L=1$, $\eta=0.5$ and $c=0.01$.
  We consider $n=2$
  branches, and show only the region of the
  circle with angular coordinate $\theta \in [0,
    \pi/2]$. The red area denoted as $l$ is the region of points reaching the
    origin without the use of the fast layer. The black region,
    labeled as $g_2$,
    corresponds to points that use the fast layer from $0 < r_\times <
    L/n$. The white dashed curve is given by
    Eq.~\ref{eq:theta_l_final}.
    Finally, the orange region, namely $g_1$, indicates the part of the system
   that use the entire length of the fast layer. The dotted white
   curve is given by Eq.~\ref{eq:theta_g_final}.
   The full white region is given by Eq.~\ref{eq:theta_g2_final}. (b) Same as in (a)
   but for $n=3$. (c) Same as in (a)
   but for $\eta=0.2$ and $c=0.1$. (d) Same as in (c)
   but for $n=3$.}
  \label{fig:S1}
\end{figure}

\subsubsection*{Boundaries between the various regions}

In the following, we determine the equations that define the boundaries between the various regions of integration of Eq.~(\ref{eq:tglo_final}).
With reference to Fig.~\ref{fig:S1}, we define three distinct parts of
the section of the circle: (i) the region $l$ composed of points
whose minimum-cost path to
the center is entirely on the slow layer; (ii) the region $g_1$
composed of points that reach the center using entire length of the fast
layer; (iii) the region $g_2$
composed of points that reach the center using part of the fast
layer.

%\subsubsection*{Boundary between the $l$ and $g_2$ regions}

The distinction between the $l$ and $g_2$ regions is determined by the
condition $q_l = q_{g_2}$, thus
\begin{equation}
  \eta  \cos \theta_{l, g_2} + \sqrt{1- \eta^2} \sin \theta_{l, g_2}
  = \frac{r - 2 c}{r}\; .
 \label{eq:theta_l}
  \end{equation}

%To solve Eq.~(\ref{eq:theta_l}), we remind that the generic equation
%  \[
%    a \sin \theta + b \cos \theta = c
%  \]
%  can be rewritten as
%  \[
%    \sqrt{a^2 + b^2} \, \sin (\theta+ \beta) = c \; , 
%  \]
%  where
%  \[
%    \beta = \arccos \frac{a}{\sqrt{a^2+b^2}} = \arcsin
%    \frac{b}{\sqrt{a^2+b^2}} \; ,
%  \]
%  thus
%  \[
%    \theta = \arcsin \frac{c}{\sqrt{a^2+b^2}} - \arcsin
%    \frac{b}{\sqrt{a^2+b^2}} \; .
%    \]
%    In the case of Eq.~\ref{eq:theta_l}, we have
%    $c =  (r- 2c)/r$, $b = \eta$ and $a= \sqrt{1-\eta^2}$, therefore we get
%    $\sqrt{a^2+b^2} = \sqrt{\eta^2 + 1  - \eta^2} = 1$, thus

After some calculations, we get
    \begin{equation}
 \theta_{l, g_2} = \arcsin \frac{r - 2 c}{r}  -  \arcsin  \eta \; .
 \label{eq:theta_l_final}
\end{equation}

%\subsubsection*{Boundary between the $g_1$ and $g_2$ regions}

The distinction between the $g_1$ and $g_2$ regions is determined by the
condition $r_\times = \ell$, thus
\begin{equation}
- \cos \theta_{g_1, g_2} + \frac{\eta}{\sqrt{1-\eta^2}} \sin \theta_{g_1, g_2} = - \ell/r \; ,
 \label{eq:theta_g}
\end{equation}
%Using the same technique as above, we have that
%$a =  \frac{\eta}{\sqrt{1-\eta^2}}$, $b = - 1$, and $c = - \ell/r$, thus
%$\sqrt{a^2 + b^2} = \sqrt{ (\eta^2 + 1 -\eta^2)/(1-\eta^2)} = 1/\sqrt{1
%  - \eta^2}$. 
%  We can therefore write
from which
\begin{equation}
 \theta_{g_1, g_2} = - \arcsin \frac{\ell \, \sqrt{1-\eta^2}}{r}  +  \arcsin  \sqrt{1-\eta^2} \; .
 \label{eq:theta_g_final}
  \end{equation}

%\subsubsection*{Boundary between the $l$ and $g_1$ regions}

The distinction between the $l$ and $g_1$ regions is determined by the
condition $r = q_{g_1}$, thus
\begin{equation}
r = \sqrt{\ell^2 + r^2 - 2 r \ell \cos \theta_{l,g_2}} + \eta \ell + 2 c
\; .
 \label{eq:theta_g2}
\end{equation}

%We can write
%\[
%  r - \eta \ell - 2 c = \sqrt{\ell^2 + r^2 - 2 r \ell \cos \theta_{l,g_2}}
%\]
%valid only if $r \geq \eta \ell + 2 c$. We have
%\[
%  (r - \eta \ell - 2 c)^2 = \ell^2 + r^2 - 2 r \ell \cos \theta_{l,g_2}
%\]
%\[
%  r^2 - 2 r (\eta \ell + 2 c) + (\eta \ell + 2 c)^2  = \ell^2 + r^2 - 2 r \ell \cos  \theta_{l,g_2}
%\]
%\[
%  2 r (\ell \cos \theta_{l,g_2} - \eta \ell - 2 c)  = \ell^2
%  - (\eta
%  \ell + 2 c)^2 
%\]
%\[
%  \ell \cos \theta_{l,g_2} - \eta \ell - 2 c  = \frac{\ell^2
%  - (\eta
%  \ell + 2 c)^2 }{2r} 
%\]
%\[
%  \ell \cos \theta_{l,g_2}  = \frac{\ell^2
%  - (\eta
%  \ell + 2 c)^2 }{2r}  +  \eta \ell +  2 c
%\]
%\[
%  \cos \theta_{l,g_2}  = \frac{1}{\ell} \left( \frac{\ell^2
%  - (\eta
%  \ell + 2 c)^2 }{2r}  +  \eta \ell +  2 c \right) \;. 
%\]
%Since $\arccos x = \pi/2 - \arcsin x$, 
%we can write
After some calculations, we get
\begin{equation}
 \theta_{l, g_1} = \frac{\pi}{2} - \arcsin   \frac{1}{\ell} \left( \frac{\ell^2
  - (\eta
  \ell + 2 c)^2 }{2r}  +  \eta \ell +  2 c \right)  \; .
 \label{eq:theta_g2_final}
\end{equation}

\newpage

\section{Real cities}

\begin{figure*}[!htb]
\includegraphics[width=0.8\textwidth]{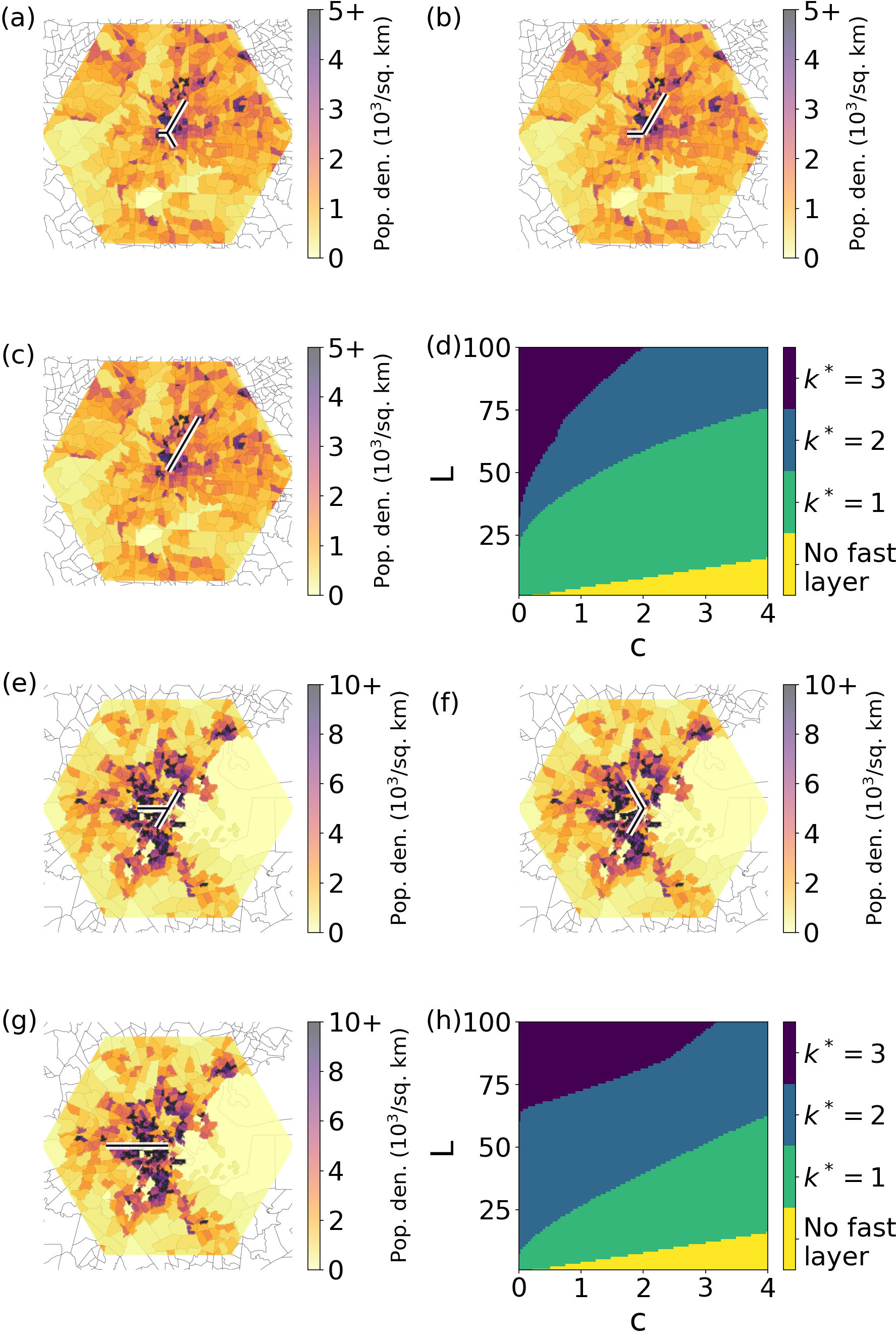}
\caption{\textbf{Phase transition in the optimal fast layer for Atlanta and Boston.} We set the city radius $\tilde{R}=20$ km for Boston and $\tilde{R}=25$ km for Atlanta, and we use a triangular lattice with radius $R=100$ to model the layers of the multiplex transportation system. 
The color map displays the population density in the city.
We show the distinct solutions obtained for the length of the fast layer $L=50$ for various values of the switching cost $c$ and $\eta =0.5$. (a) Optimal fast network with the degree of center $k^*=1$ for Atlanta. (b) Same as in (a) with $k^*=2$. (c) Same as in (a) with $k^*=1$. (d) We show the heat map illustrating the type of optimal solution (shown in panels (a) and (b)) obtained for different combinations of $c$ and $L$. 
%We vary the values of $c$ and $L$ from $0$ to $4$ and $1$ to $100$, respectively.
The yellow region indicates that building a fast layer for the values of $c$ and $L$ does affect the 
%shortest
minimum-cost
paths to the center regardless of the structure, i.e., $L<r_c$. (e) Optimal fast network with the degree of center $k^*=1$ for Boston. (f) Same as in (e) with $k^*=2$. (f) Same as in (e) with $k^*=1$. (h) Same as in (d) for Boston.}
  \label{fig:main2}
\end{figure*}

\begin{figure*}[!htb]
\includegraphics[width=\textwidth]{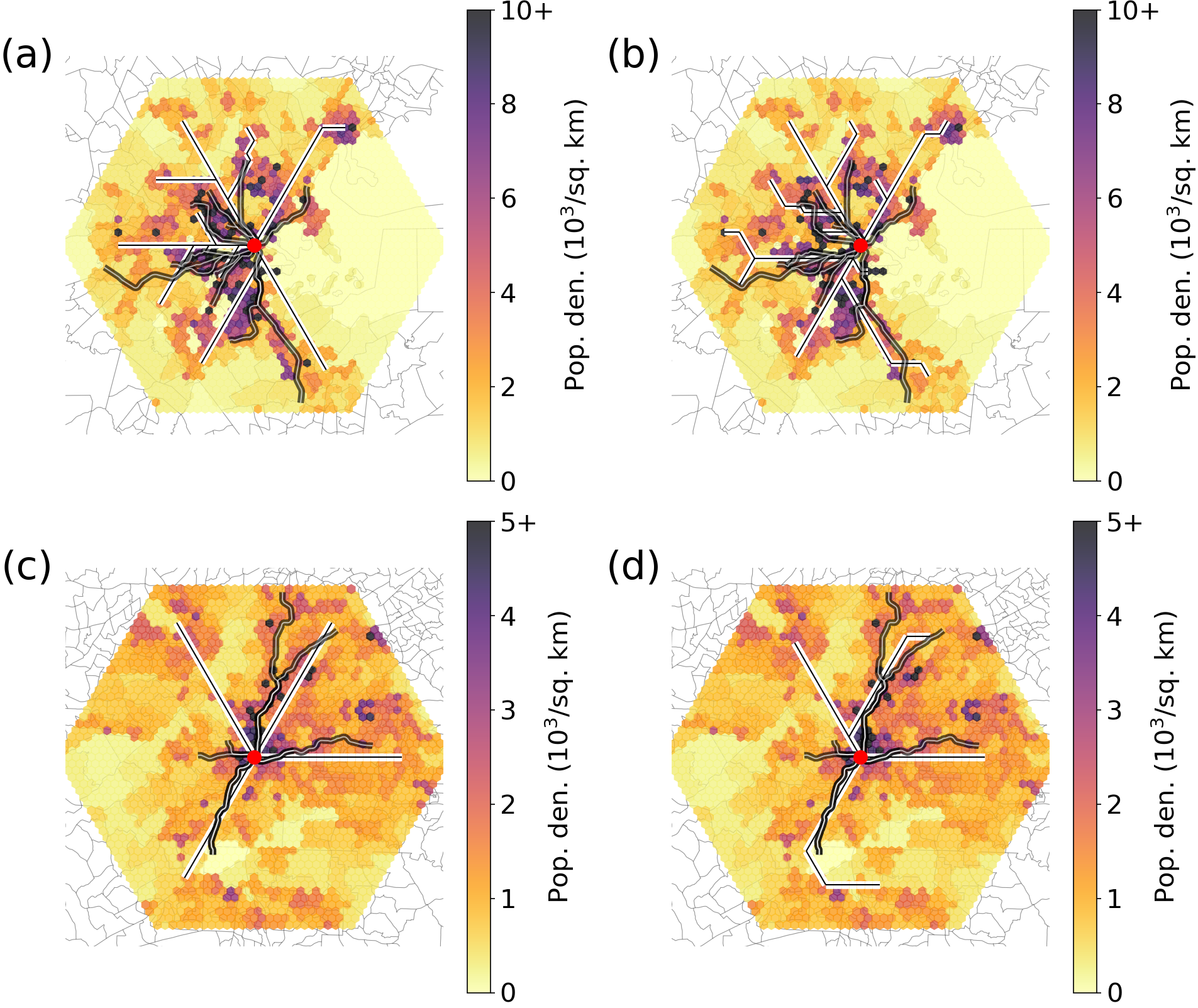}
\caption{\textbf{Illustration of the optimal fast layer for Atlanta and Boston.} %We construct the slow layer of the system using a triangular lattice with a city radius $\tilde{R}=20$ km for Boston and $\tilde{R}=25$ km for Atlanta with lattice radius $R=25$ for both. The color map shows the population density associated with lattice points in $10^3$ individuals per $km^2$. To show the heterogeneity in the population density we cap the color maps to $5000$ and $10000$ in Atlanta and Boston. The thick white lines with thin black lines show the solutions obtained from the greedy algorithm for switching cost $c=3$ minutes and fast layer speed $40$ km/h. The cyan circle denotes the center of the system. 
We compare the solution obtained from our optimization algorithm (thick white lines) with the real subway system (thick black curves) in Boston (a, b) Atlanta (c, d) for two different parameter settings corresponding to low (a, c) and high (b, d) congestion. 
(a) The optimized configuration for Boston is obtained by setting $\eta = 0.125$ and $c = 1.25$, corresponding to a switching time of $3$ minutes and slow and fast layer speeds $20$ km/h and $40$ km/h, respectively. (b) same as (a) but for $\eta = 0.5$ and $c = 0.3125$, corresponding to a switching time of $3$ minutes and slow and fast layer speeds $5$ km/h and $40$ km/h, respectively. (c) The optimized configuration for Atlanta for low congestion is obtained by setting $\eta = 0.125$ and $c = 1$, corresponding to a switching time of $3$ minutes and slow and fast layer speeds $20$ km/h and $40$ km/h, respectively. (d) same as (c) but for $\eta = 0.125$ and $c = 0.25$, corresponding to a switching time of $3$ minutes and slow and fast layer speeds $5$ km/h and $40$ km/h, respectively.
The heat maps are used to represent population densities in the cities.
}
\label{fig:main3}
\end{figure*}

In the main paper, we show the phase transition between different types of solutions in Toronto. In Figures~\ref{fig:main2} and~\ref{fig:main3}, we display results of a similar analysis for Boston and Atlanta.

\end{document}